\documentclass[
    pre,
    aps,
    amsmath,
    amssymb,
    reprint,
    superscriptaddress,
    nofootinbib
]{revtex4-2}

\usepackage{graphicx}
\usepackage{dcolumn}
\usepackage{bm}
\usepackage[
    bookmarks=false,
    colorlinks=true,
    linkcolor=blue,
    citecolor=blue,
    urlcolor=blue,
]{hyperref}
\usepackage{orcidlink}
\usepackage{titlesec}

\definecolor{aesthetic-background}{RGB}{30, 20, 40}
\definecolor{aesthetic-blue}{RGB}{0, 174, 255}
\definecolor{aesthetic-cyan}{RGB}{90, 200, 230}
\definecolor{aesthetic-green}{RGB}{55, 196, 55}
\definecolor{aesthetic-magenta}{RGB}{249, 42, 173}
\definecolor{aesthetic-yellow}{RGB}{253, 163, 42}
\definecolor{ppt-blue}{RGB}{2, 83, 118}

\makeatletter
\newcommand*\@secondofsix[6]{#2}
\newcommand{\addtotitleformat}{%
  \@ifstar{\addtotitleformat@star}{\addtotitleformat@nostar}}
\newcommand\addtotitleformat@nostar[2]{%
  \PackageError{titlesec}{non starred form of \string\addtotitleformat\space not supported}{}}
\newcommand\addtotitleformat@star[2]{%
  \expandafter\expandafter\expandafter\expandafter
  \expandafter\expandafter\expandafter\def
  \expandafter\expandafter\expandafter\expandafter
  \expandafter\expandafter\expandafter\@currentsection@font
  \expandafter\expandafter\expandafter\expandafter
  \expandafter\expandafter\expandafter{%
    \expandafter\expandafter\expandafter\@secondofsix
       \csname ttlf@\expandafter\@gobble\string#1\endcsname}%
  \titleformat*{#1}{\@currentsection@font#2}%
}
\makeatother

\addtotitleformat*{\section}{\MakeUppercase}

\makeatletter
\renewcommand*\date[1][\Dated@name]{
    \def\@date{
        #1\today
    }
}
\makeatother

\newcommand{\tiiint}{\int\!\!\!\!\!\!\:\int\!\!\!\!\!\!\:\int}

\begin{document}


\title{
%
    Modeling single-molecule stretching experiments using statistical thermodynamics
}
\author{
    Michael R. Buche%
    \:\orcidlink{0000-0003-1892-0502}\,
}
\email{mrbuche@sandia.gov}
\affiliation{
    Computational Solid Mechanics and Structural Dynamics, Sandia National Laboratories, Albuquerque, New Mexico 87185, USA
}
\author{
    Jessica M. Rimsza%
    \:\orcidlink{0000-0003-0492-852X}\,
}
\affiliation{
    Geochemistry, Sandia National Laboratories, Albuquerque, New Mexico 87185, USA
}
\date{}

\begin{abstract}
Single-molecule stretching experiments are widely utilized within the fields of physics and chemistry to characterize the mechanics of individual bonds or molecules, as well as chemical reactions.
Analytic relations describing these experiments are valuable, and these relations can be obtained through the statistical thermodynamics of idealized model systems representing the experiments.
Since the specific thermodynamic ensembles manifested by the experiments affect the outcome, primarily for small molecules, the stretching device must be included in the idealized model system.
Though the model for the stretched molecule might be exactly solvable, including the device in the model often prevents analytic solutions.
In the limit of large or small device stiffness, the isometric or isotensional ensembles can provide effective approximations, but the device effects are missing.
Here, a dual set of asymptotically correct statistical thermodynamic theories are applied to develop accurate approximations for the full model system that includes both the molecule and the device.
The asymptotic theories are first demonstrated to be accurate using the freely jointed chain model, and then using molecular dynamics calculations of a single polyethylene chain.
\smallskip\smallskip\smallskip

\noindent DOI: \href{https://doi.org/10.1103/PhysRevE.108.064503}{10.1103/PhysRevE.108.064503}
\end{abstract}

\maketitle


\section{Introduction}\label{sec:introduction}

The applications of mechanical forces at the molecular level is an effective method of understanding molecular mechanisms, as well as the underlying chemical physics.
Single-molecule stretching experiments examining these mechanisms have been made possible using devices like atomic force microscopes \cite{binnig1986atomic,meyer1988novel,florin1994adhesion,lee1994direct,ortiz1999entropic,giessibl2003advances,dudko2008theory,abkenar2017dissociation} or optical tweezers \nolinebreak \cite{ashkin1970acceleration,ashkin1970atomic,chu1985three,chu1986experimental,ashkin1987optical,grier2003revolution,de2015temperature,gieseler2021optical,volpe2023roadmap}.
The complex folding mechanisms and related biomolecular functions of proteins, nucleic acids, and other biological macromolecules have been studied via single-molecule stretching experiments with devices like these \nolinebreak \cite{smith1996overstretching,mehta1999single,woodside2006direct,woodside2014reconstructing,dutta2016probing,bustamante2020single,bustamante2021optical}.
Utilizing mechanochemistry \cite{garcia2017steering,do2017mechanochemistry,o2021many}, reactive molecules known as mechanophores are often incorporated into polymer materials, in order to activate desired chemical functionality when the material is deformed \nolinebreak \cite{black2011molecular,brantley2013polymer,vidavsky2019enabling,chen2021mechanochemical,jayathilaka2021force,ghanem2021role}.
Similar devices used in single-molecule stretching experiments for biological systems are also used to characterize mechanophores in order to assist material design \nolinebreak \cite{evans2001probing,liang2009mechanochemistry,wang2015inducing,gossweiler2015force,zhang2017multi,akbulatov2017experimental,barbee2018substituent,sulkanen2019spatially,liu2020polymer,stratigaki2020methods}.
Since mathematical modeling is commonly used to support all of these experiments, it is clearly a worthwhile endeavor to improve the associated models.

Many of the existing models that correspond to single-molecule stretching experiments utilize the principles of statistical thermodynamics, since thermal effects are often important if not dominant (especially for polymers).
Simulations techniques such as density functional theory and molecular dynamics can incorporate or approximate thermal effects in modeling molecular stretching \nolinebreak \cite{moses2004characterization,huang2017single,vidavsky2020tuning,hsissou2021synthesis}, but this is typically expensive, and further does not produce interpretable relations for practical modeling goals.
Analytic relations for the experiments are valuable since they enable rapid prediction and further understanding.
To that end, a common modeling approach involves constructing an idealized model system that represents the experiment, applying the principles of statistical thermodynamics (i.e. formulating the partition function), and analytically solving the system for the desired relations.
Since an experiment creates a specific thermodynamic ensemble for the system (and since different thermodynamic ensembles generally result in different outcomes, especially for things like short polymer chains \nolinebreak \cite{neumann2003precise,sinha2005inequivalence,suzen2009ensemble,manca2014equivalence,dutta2018inequivalence,benedito2018isotensional,buche2020statistical}) the stretching device should be included in the model.
This is done by including a device potential in the model system Hamiltonian, and in limited cases the resulting partition functions can be integrated analytically \nolinebreak \cite{benedito2020rate,florio2019unveiling,bellino2019influence}.
Unfortunately, models in general -- even models which are analytically solvable before including the device -- become unsolvable once including the device within the model.
If the device stiffness is sufficiently large or small, it can be accurate to approximate the model system using the isometric or isotensional thermodynamic ensembles \nolinebreak \cite{kreuzer2001stretching,florio2019unveiling,bellino2019influence}, but the effects of the device are completely lost.

In order to obtain accurate analytic relations for arbitrary molecules under the effects of the device, a dual set of asymptotically correct statistical thermodynamic theories are developed here starting from the general theory.
The first theory builds approximations for stiff devices using the asymptotic theory for systems with steep potentials \cite{buche2021fundamental,buche2021chain,buche2022freely,mulderrig2023statistical,buche2024statistical}.
The second theory builds approximations for compliant devices using Zwanzig's theory for systems with weak potentials \cite{zwanzig1954high,mcq}.
The accuracy of both theories is verified by applying them to the freely jointed chain model and molecular dynamics calculations of a single polyethylene chain.
Overall, this work provides a rigorous and systematic approach of including a device when modeling single-molecule stretching experiments.

\clearpage

\section{General theory}\label{sec:theory}

A modified rendition of the canonical ensemble \cite{mcq} is utilized in order to model the finite stiffness of the device stretching the molecule.
One end of the molecule is fixed at the origin, while the other end is subjected to a harmonic potential of stiffness $W$ located at $\boldsymbol{\xi}$.
The temperature $T$ is fixed and parameterized by $\beta=1/kT$, where $k$ is the Boltzmann constant.
Only configurational degrees of freedom are considered without loss of generality.
The partition function for the system $Q$ can be calculated by integrating the partition function for the molecule at a fixed end-to-end vector (the isometric ensemble) over all end-to-end vectors $\boldsymbol{\xi}'$, weighted by the Boltzmann factor resulting from the device potential centered at $\boldsymbol{\xi}$.
This modified canonical ensemble partition function $Q$ is then

\begin{equation}\label{eq:Q-exact}
    Q(\boldsymbol{\xi}) =
    \tiiint Q_0(\boldsymbol{\xi}') \, e^{-\tfrac{1}{2}\beta W\left(\boldsymbol{\xi} - \boldsymbol{\xi}'\right)^2} d^3\boldsymbol{\xi}'
    ,
\end{equation}
where $Q_0(\boldsymbol{\xi}')$ is the partition function of the system in the isometric ensemble, where the end-to-end vector $\boldsymbol{\xi}'$ of the molecule would be fixed.
The expected force $\mathbf{f}$ is an ensemble average can then be calculated from $Q$ as

\begin{equation}\label{eq:f-exact}
    \mathbf{f}(\boldsymbol{\xi}) =
    \frac{-1\phantom{-}}{\beta}\frac{\partial\ln Q(\boldsymbol{\xi})}{\partial\boldsymbol{\xi}}
    .
\end{equation}
To similarly calculate the expected end-to-end vector of the molecule, it becomes convenient to define another partition function $Z\equiv e^{\beta W\xi^2/2} Q$ and a force effectively applied by the potential $\mathbf{f}\equiv W\boldsymbol{\xi}$.
This results in

\begin{equation}\label{eq:Z-exact}
    Z(\mathbf{f}) =
    \tiiint Q_0(\boldsymbol{\xi}') \, e^{\beta \mathbf{f}\cdot\boldsymbol{\xi}'} e^{-\tfrac{1}{2}\beta W\left(\boldsymbol{\xi}'\right)^2} d^3\boldsymbol{\xi}'
    ,
\end{equation}
which then allows the expected end-to-end vector $\boldsymbol{\xi}$, another ensemble average, to be calculated from $Z$ as

\begin{equation}\label{eq:xi-exact}
    \boldsymbol{\xi}(\mathbf{f}) =
    \frac{1}{\beta}\frac{\partial\ln Z(\mathbf{f})}{\partial\mathbf{f}}
    .
\end{equation}
In general, Eqs.~\eqref{eq:f-exact} and \eqref{eq:xi-exact} will produce different results.

\section{Asymptotic theory}\label{sec:asymptotic}

A duel set of asymptotically correct statistical thermodynamic theories are now applied to the general theory.
Namely, the asymptotic theory of \citet{buche2021fundamental} is used to approximate Eqs.~\eqref{eq:Q-exact} and \eqref{eq:f-exact} in the limit of a strong potential, and the perturbation theory of \citet{zwanzig1954high} is used to approximate Eqs.~\eqref{eq:Z-exact} and \eqref{eq:xi-exact} in the limit of a weak potential.
Either approach builds upon a reference system (the isometric and isotensional ensembles, respectively) and includes small corrections.
Derivatives of functions with respect to the argument are denoted using apostrophes, and vector arguments are replaced by scalars without loss of generality.

\vfill\eject

\subsection{Strong potential}\label{sec:asymptotic.strong}

As $\beta W$ becomes sufficiently large, the asymptotic theory of \citet{buche2021fundamental} shows that Eq.~\eqref{eq:Q-exact} can be asymptotically approximated as

\begin{equation}\label{eq:Q-asymptotic}
    Q(\xi) \sim
    \left(\frac{2\pi}{\beta W}\right)^{3/2} Q_0(\xi) \left[1 + \frac{1}{2\beta W}\frac{Q_0''(\xi)}{Q_0(\xi)}\right]
    ,
\end{equation}
where $Q_0(\xi)$ is the partition function of the reference system, i.e., the system in the isometric ensemble.
If $\beta W$ is indeed large, one can further approximate

\begin{equation}
    \ln\left[1 + \frac{1}{2\beta W}\frac{Q_0''(\xi)}{Q_0(\xi)}\right] \sim
    \frac{1}{2\beta W}\frac{Q_0''(\xi)}{Q_0(\xi)}
    .
\end{equation}
Using $Q_0''/Q_0=(\beta f_0)^2 - \beta f_0'$ and applying Eq.~\eqref{eq:f-exact} then yields an asymptotic relation for the expected force $f$ as a function of the applied potential distance $\xi$,

\begin{equation}\label{eq:f-asymptotic}
    f(\xi) \sim
    f_0(\xi) - \frac{1}{\beta W}\left[\beta f_0(\xi)f_0'(\xi) - \frac{f_0''(\xi)}{2}\right]
    ,
\end{equation}
valid as $\beta W$ becomes sufficiently large, where $f_0(\xi)$ is the expected force $f_0$ as a function of the applied end-to-end length $\xi$ for the system in the isometric ensemble.

\subsection{Weak potential}\label{sec:asymptotic.weak}

As $\beta W$ becomes sufficiently small, the perturbation theory of \citet{zwanzig1954high} shows that Eq.~\eqref{eq:Z-exact}, in this case, can be asymptotically approximated as

\begin{equation}\label{eq:Z-asymptotic}
    Z(f) \sim
    Z_0(f) \left[1 - \frac{W}{2\beta}\frac{Z_0''(f)}{Z_0(f)}\right]
    ,
\end{equation}
where $Z_0(f)$ is the partition function of the reference system, i.e., the system in the isotensional ensemble.
Note that $Z_0''/Z_0=(\beta\xi_0)^2$ has been used in obtaining Eq.~\eqref{eq:Z-asymptotic}, which with the assumption that $\beta W$ is small also permits 

\begin{equation}
    -\ln\left[1 - \frac{\beta W}{2}\,\xi_0^2(f)\right] \sim
    \frac{\beta W}{2}\,\xi_0^2(f)
    .
\end{equation}
Applying Eq.~\eqref{eq:xi-exact} then yields an asymptotic relation for the expected end-to-end length $\xi$ as a function of the effective force applied by the potential $f$,

\begin{equation}\label{eq:xi-asymptotic}
    \xi(f) \sim
    \xi_0(f) \Big[1 - W \xi_0'(f)\Big]
    ,
\end{equation}
valid as $\beta W$ becomes sufficiently small, where $\xi_0(f)$ is the expected end-to-end length $\xi_0$ as a function of the applied force $f$ for the system in the isotensional ensemble.

\clearpage

\section{Application: FJC model}\label{sec:fjc}

The dual set of asymptotic relations developed in Sec.~\ref{sec:asymptotic} are now demonstrated using the freely jointed chain (FJC) model.
The freely jointed chain model consists of $N_b$ rigid links of length $\ell_b$, which can pass through one another and rotate about the connecting hinges without penalty \cite{treloar1949physics,flory1969statistical,rubinstein2003polymer}.
The case of an increasingly strong applied potential is considered first, followed by the case of an increasingly weak applied potential.
In either case, the relevant asymptotic relation performs increasingly well.
All numerical calculations in this section were completed using the \texttt{Polymers Modeling Library} \cite{polymers}.

The full system is nondimensionalized using the nondimensional force $\eta\equiv\beta f\ell_b$, the nondimensional potential distance or end-to-end length per link $\gamma\equiv\xi/N_b\ell_b$, and the nondimensional potential stiffness $\varpi\equiv\beta W\ell_b^2$.
Applying these nondimensional variables to Eqs.~\eqref{eq:Q-exact} and \eqref{eq:f-exact}, the exact relation for the expected nondimensional force $\eta$ as a function of the applied nondimensional potential distance $\gamma$ is given by

\begin{equation}\label{eq:eta-fjc-exact}
    \eta(\gamma) =
    \frac{-1\phantom{-}}{N_b}\frac{\partial}{\partial\gamma}\,\ln\tiiint Q_0(\boldsymbol{\gamma}') \, e^{-\tfrac{\varpi}{2} N_b^2\left(\boldsymbol{\gamma} - \boldsymbol{\gamma}'\right)^2} d^3\boldsymbol{\gamma}'
    .
\end{equation}
Similarly applying the same nondimensional variables to Eqs.~\eqref{eq:Z-exact} and \eqref{eq:xi-exact}, the exact relations for the expected nondimensional end-to-end length per link $\gamma$ as a function of the effective applied nondimensional force $\eta$ is given by

\begin{equation}\label{eq:gamma-fjc-exact}
    \gamma(\eta) =
    \frac{1}{N_b}\frac{\partial}{\partial\eta}\,\ln \tiiint Q_0(\boldsymbol{\gamma}') \, e^{N_b\boldsymbol{\gamma}'\cdot\big(\boldsymbol{\eta} - \tfrac{N_b\varpi}{2}\boldsymbol{\gamma}'\big)} d^3\boldsymbol{\gamma}'
    .
\end{equation}
The integrals in Eqs.~\eqref{eq:eta-fjc-exact} and \eqref{eq:gamma-fjc-exact} can each be reduced to one dimension, subsequently integrated numerically.

\subsection{Strong potential}\label{sec:fjc.strong}

For a large nondimensional potential stiffness $\varpi\gg 1$, the expected nondimensional force $\eta$ as a function of the applied nondimensional potential distance $\gamma$ given by Eq.~\eqref{eq:eta-fjc-exact} can instead be asymptotically approximated to $\mathrm{ord}(\varpi^{-1})$ using Eq.~\eqref{eq:f-asymptotic} as

\begin{equation}\label{eq:eta-fjc-asymptotic}
    \eta(\gamma) \sim
    \eta_0(\gamma)
    - \frac{1}{N_b\varpi}\left[\eta_0(\gamma)\eta_0'(\gamma) - \frac{\eta_0''(\gamma)}{2N_b}\right]
    ,
\end{equation}
where $\eta_0$ is the expected nondimensional force as a function of an applied nondimensional end-to-end length per link $\gamma$ for a freely jointed chain in the isometric ensemble,

\begin{equation}
    \eta_0(\gamma) =
    \frac{1}{N_b\gamma} + \left(\frac{1}{2} - \frac{1}{N_b}\right)\frac{h(\gamma, 3)}{h(\gamma, 2)}
    ,
\end{equation}
which has been analytically calculated here using the exact equilibrium distribution $P^\mathrm{eq}$ given by \citet{treloar1949physics} and

\begin{figure}[t]
    \begin{center}
        \includegraphics{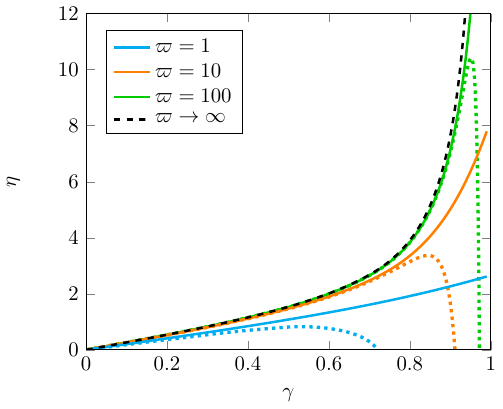}
    \end{center}
    \caption{\label{fig:fjc-strong}%
        (FJC model) Expected nondimensional force $\eta$ as a function of the applied nondimensional potential distance $\gamma$, calculated using the asymptotic (dotted) and exact (solid) approaches, for varying nondimensional potential stiffness $\varpi$.
    }
\end{figure}

\vfill\eject

\noindent
the pertinent thermodynamic relation \cite{buche2020statistical}

\begin{equation}
    \begin{aligned}
        P^\mathrm{eq}(\gamma) = & \
        \frac{(N_b\ell_b)^3 N_b^{N_b}}{8\pi (N_b - 2)!}\frac{h(\gamma, 2)}{\gamma}
        ,\\
        \eta(\gamma) = & \
        \frac{1}{N_b}\frac{\partial}{\partial\gamma}\,\ln P^\mathrm{eq}(\gamma)
        .
    \end{aligned}
\end{equation}
The auxiliary functions $h(\gamma,n)$, where the summands are given by $s_\mathrm{max}(\gamma)\equiv\lfloor(1-\gamma/2)N_b\rfloor$, are defined as

\begin{equation}
    h(\gamma, n) \equiv
    \sum_{s=0}^{s_\mathrm{max}(\gamma)}(-1)^s\binom{N_b}{s}\left(\frac{1-\gamma}{2} - \frac{s}{N_b}\right)^{N_b - n}
    .
\end{equation}
The exact relation in Eq.~\eqref{eq:eta-fjc-exact} is plotted with the asymptotic relation in Eq.~\eqref{eq:eta-fjc-asymptotic} in Fig.~\ref{fig:fjc-strong}.
As $\varpi$ increases, the asymptotic approach does increasingly well in approximating the full system, but tends to fail as $\gamma\to 1$.
This eventual failure is because the freely jointed chain model experiences increasingly large forces as $\gamma\to 1$ due to its inextensibilty [roughly, $\eta\sim(1-\gamma)^{-1}$ as $\gamma\to 1$].
Once the nondimensional force $\eta$ starts to near the order of the nondimensional potential stiffness $\varpi$, the end of the chain will be necessarily be biased away from the center of the potential well.
This in turn causes the asymptotic approach, which in this case is based on the isometric ensemble, to become inaccurate.
It is then apparent that $\varpi\gg\eta$ and $\varpi\gg 1$ are both requirements for the asymptotic approach to succeed in accurately approximating the full system.
Therefore, experiments that attempt to apply isometric conditions will achieve higher accuracy at larger extensions with molecules that are less stiff.

\clearpage

\subsection{Weak potential}\label{sec:fjc.weak}

\begin{figure}[t]
    \begin{center}
        \includegraphics{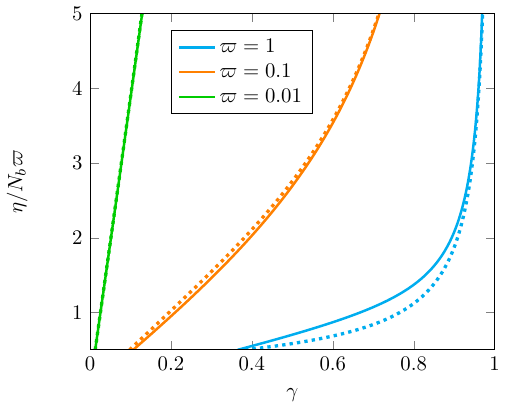}
    \end{center}
    \caption{\label{fig:fjc-weak}%
        (FJC model) Rescaled effective applied nondimensional force $\eta/N_b\varpi$ as a function of the expected nondimensional end-to-end length per link $\gamma$, calculated using the asymptotic (dotted) and exact (solid) approaches, for varying nondimensional potential stiffness $\varpi$.
    }
\end{figure}

For a small nondimensional potential stiffness $\varpi\ll 1$, the expected nondimensional end-to-end length per link $\gamma$ as a function of the effective applied nondimensional force $\eta$ given by Eq.~\eqref{eq:gamma-fjc-exact} can instead be asymptotically approximated to $\mathrm{ord}(\varpi)$ using Eq.~\eqref{eq:xi-asymptotic} as

\begin{equation}\label{eq:gamma-fjc-asymptotic}
    \gamma(\eta) \sim
    \gamma_0(\eta)\Big[1 - N_b\varpi\gamma_0'(\eta)\Big]
    ,
\end{equation}
where $\gamma_0$ is the expected nondimensional end-to-end length per link as a function of an applied nondimensional force $\eta$ for a freely jointed chain in the isotensional ensemble, given exactly by the Langevin function \cite{rubinstein2003polymer}

\begin{equation}
    \gamma_0(\eta) =
    \coth(\eta) - \frac{1}{\eta}
    .
\end{equation}
The exact relation in Eq.~\eqref{eq:gamma-fjc-exact} is plotted with the asymptotic relation in Eq.~\eqref{eq:gamma-fjc-asymptotic} in Fig.~\ref{fig:fjc-weak}.
As $\varpi$ decreases and/or the nondimensional force $\eta$ increases (the applied nondimensional potential distance $\eta/N_b\varpi$ increases), the asymptotic approach does increasingly well in approximating the full system.
Notably, the asymptotic approach appears to succeed for sufficiently distant potentials for any value of $\varpi$.
This is an artifact resulting from the inextensibility of the freely jointed chain model, where even a stiff potential ($\varpi\gg 1$) that is distant ($\eta/N_b\varpi\gg 1$) will not stretch the chain past $\gamma=1$.
For more general molecular models, the effective molecular stiffness would compete with the potential stiffness, such that the potential would need to be weak ($\varpi\ll 1$) in addition to distant ($\eta/N_b\varpi\gg 1$) in order for the asymptotic approach to become accurate.
Otherwise, the distant potential would be strong enough to stretch the molecule to lengths that compare with the potential distance, which would cause appreciable fluctuations in the resulting force.
Therefore, experiments that attempt to apply isotensional conditions will achieve higher accuracy at larger forces with molecules that are more stiff.

While the approximation for strong potentials in Eq.~\eqref{eq:eta-fjc-asymptotic} improves as the molecular system becomes large ($N_b$ increases), the approximation for weak potentials in Eq.~\eqref{eq:gamma-fjc-asymptotic} instead worsens.
This contrasts with common intuition concerning approximations, thermodynamic ensembles, and system size, especially when stretching polymer chains \cite{mcq,neumann2003precise,suzen2009ensemble,manca2014equivalence,buche2020statistical}.
This reversal is simply because the approximation for weak potentials, based on the isotensional ensemble, relies on the total length scale of the molecule ($N_b\ell_b$ in this case) being small compared to the distance of the potential.
The resulting disparity in length scales is what causes the fluctuations in end-to-end distance to become negligible compared to the shape of the potential, and the resulting effective force from the potential to become approximately constant.
Therefore, experiments that attempt to apply isotensional conditions will achieve higher accuracy at larger distances with molecules that are smaller in effective length.

\section{Application: Polyethylene}\label{sec:md}

\begin{figure}[b]
    \begin{center}
        \includegraphics[width=0.2\textwidth]{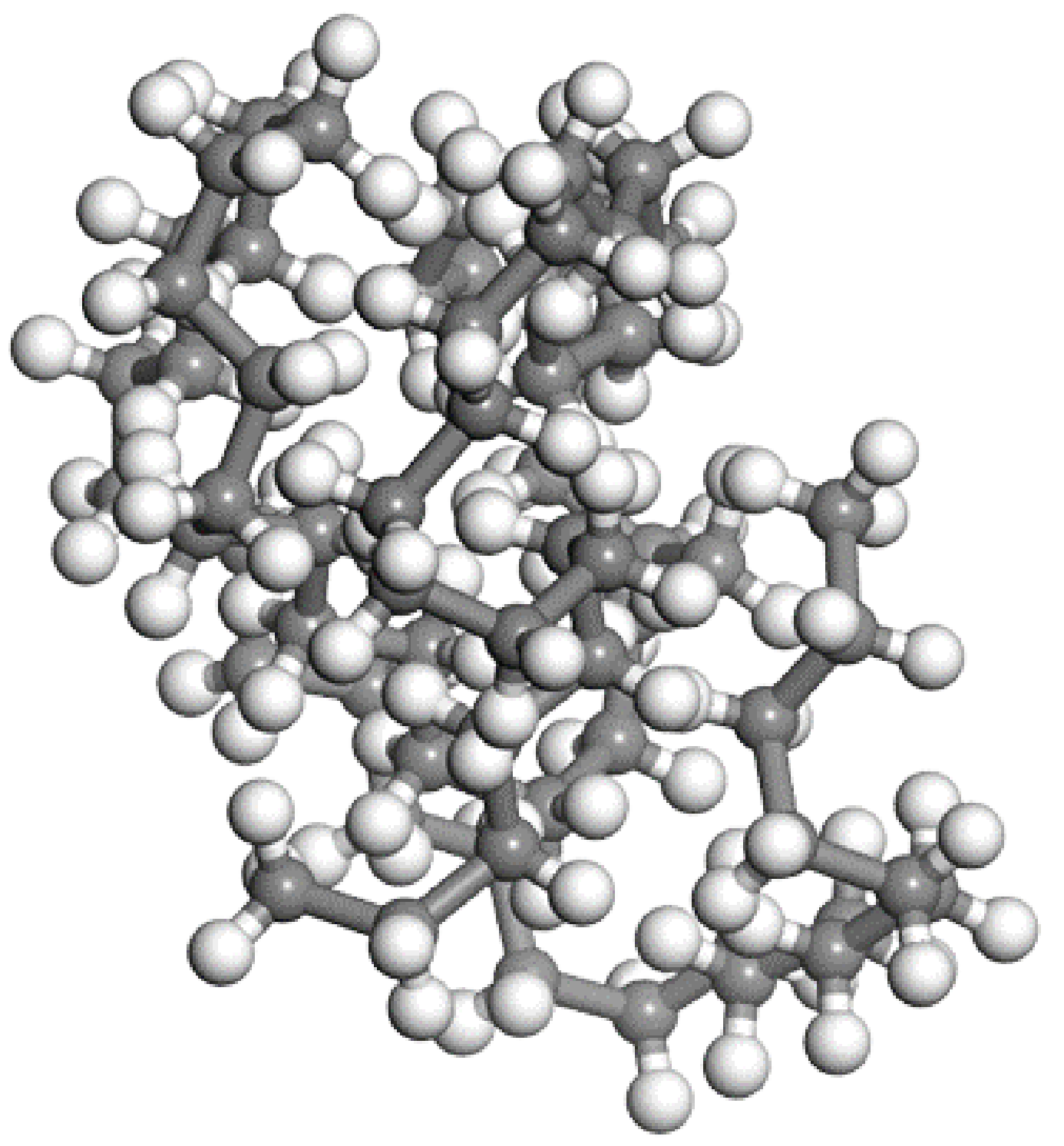}
    \end{center}
    \caption{\label{fig:md-polyethylene}%
        Short polyethylene chain after relaxation, where \\ carbon atoms are gray and hydrogen atoms are white.
    }
\end{figure}

The dual set of asymptotic relations developed in Sec.~\ref{sec:asymptotic} are now demonstrated using a molecular dynamics (MD) calculation of a single short polyethylene chain.
The model system consisted of a saturated chain of 64 carbon atoms (see Fig.~\ref{fig:md-polyethylene}) in a cube of volume 512~nm$^2$ at temperature 100~K (NVT) with a damping parameter of 1.0~fs for the thermostat.
Keeping one of the two terminal carbon atoms fixed, the end-to-end length $\xi$ was calculated at each timestep (0.1~fs, due to the possibility of fast proton dynamics) over the total time of 300~ns (totaling 3~billion samples).
These calculations were completed using \texttt{LAMMPS} \cite{lammps} with the reactive bond-order based force-field \texttt{ReaxFF} parametrized (see Supporting Information) for hydrocarbons \cite{mueller2010application}, where it was verified that no reactions occurred.

\clearpage

\subsection{Strong potential}\label{sec:md.strong}

\begin{figure}[t]
    \begin{center}
        \includegraphics{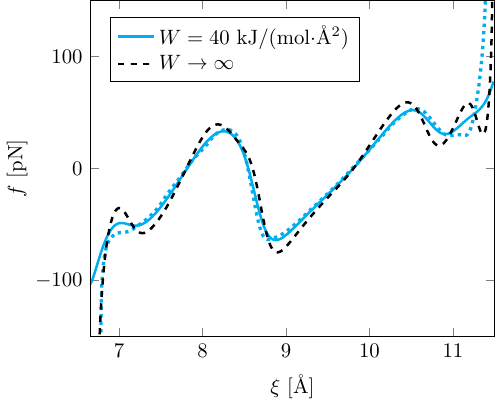}
    \end{center}
    \caption{\label{fig:md-strong}%
        (Polyethylene) Expected force $f$ as a function of the applied potential distance $\xi$, calculated using the asymptotic (dotted) and molecular dynamics (solid) approaches.
    }
\end{figure}

The equilibrium radial distribution function $g^\mathrm{eq}_0$ was generated from the MD calculations using a smoothed histogram of 1000 equal-length bins between the minimum (6.2~\AA) and maximum (11.3~\AA) observed values of the end-to-end length $\xi$.
The equilibrium distribution $P^\mathrm{eq}_0=(2\pi\xi^2)^{-1}g^\mathrm{eq}_0\propto Q_0$ was then used to obtain the expected mechanical response $f(\xi)$ through Eqs.~\eqref{eq:Q-exact} and \eqref{eq:f-exact}.
The asymptotic approximation for $f(\xi)$ was calculated using Eq.~\eqref{eq:f-asymptotic}.
The results are shown in Fig.~\ref{fig:md-strong} for a potential stiffness of $W=40$~kJ/(mol$\cdot$\AA$^2$), which if $\ell_b=1.5$~\AA\ is chosen (approximate carbon-carbon bond length), then $W$ corresponds to $\varpi\approx 100$.
The results in Fig.~\ref{fig:md-strong} show the asymptotic approach provides an accurate approximation until the forces increase rapidly and are dominated by the potential stiffness.
Figure~\ref{fig:md-strong} also shows that the asymptotic approach is applicable to molecules under compression.
Unlike the links in FJC model, the polyethylene monomers in this MD calculation interact and cannot overlap.
This causes complicated conformational changes as the molecule is extended, as evidenced in Fig.~\ref{fig:md-strong} by the nonmonotonic behavior of the force as a function of the applied extension.

\subsection{Weak potential}\label{sec:md.weak}

\begin{figure}[t]
    \begin{center}
        \includegraphics{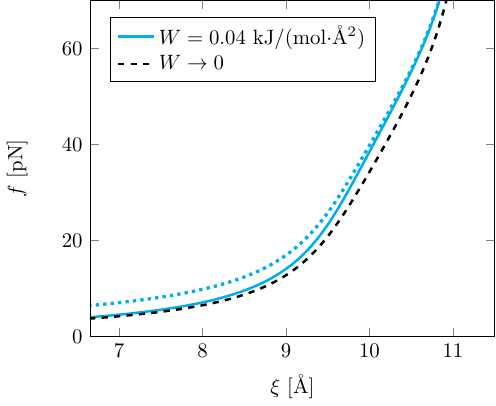}
    \end{center}
    \caption{\label{fig:md-weak}%
        (Polyethylene) Effective force $f$ as a function of the expected end-to-end length $\xi$, calculated using the asymptotic (dotted) and molecular dynamics (solid) approaches.
    }
\end{figure}

The expected end-to-end length as a function of the effective applied force $\xi(f)$, where the effective force $f$ is defined as the potential stiffness $W$ multiplying the potential distance, was calculated using $Q=e^{-\beta f^2/2W}Z$ from Sec.~\ref{sec:md.strong} and Eqs.~\eqref{eq:Z-exact} and \eqref{eq:xi-exact}.
The asymptotic approximation for $\xi(f)$ was calculated using Eq.~\eqref{eq:xi-asymptotic}.
The results are shown in Fig.~\ref{fig:md-weak} for a potential stiffness of $W=0.04$~kJ/(mol$\cdot$\AA$^2$), which if $\ell_b=1.5$~\AA\ is chosen (approximate carbon-carbon bond length), then $W$ corresponds to $\varpi\approx 0.1$.
The asymptotic approach provides an accurate approximation as the potential distance ($f/W$) increases, as predicted by the asymptotic theory.
Note that Fig.~\ref{fig:md-weak} shows purely monotonic behavior, in contrast to the nonmonotonic behavior shown in Fig.~\ref{fig:md-strong}.
For short potential distances, the weak potential provides a small effective force that does not inhibit conformational changes of the molecule.
The resulting ensemble average end-to-end length as a function of the effective force is then monotonic as the potential distance increases.
As the potential distance and corresponding effective force eventually do become large, the molecule is nearing full extension and no longer exhibits conformational changes, so the end-to-end length remains monotonic.

\section{Conclusion}\label{sec:conclusion}

The statistical thermodynamic theory for modeling single-molecule stretching experiments was presented, and dual asymptotic approximations were developed.
The accuracy of these approximations is based on the nondimensional parameter $\varpi=\beta WL^2$, where $W$ is the device stiffness and $L$ the molecular length scale.
When modeling experiments with stiff devices at short distances ($\varpi\gg 1$), the isometric ensemble provides an approximation with $O(\varpi^{-1})$ error, and the asymptotic theory building upon it provides an approximation with $O(\varpi^{-2})$ error.
When modeling experiments with compliant devices at large distances ($\varpi\ll 1$), the isotensional ensemble provides an approximation with $O(\varpi)$ error, and the asymptotic theory building upon it provides an approximation with $O(\varpi^{2})$ error.
The effectiveness of the asymptotic approach was demonstrated analytically using the freely jointed chain model, and numerically using a molecular dynamics calculation of a polyethylene chain.
These results seemed to indicate that $\varpi=100$ and $\varpi=0.01$ are reasonable thresholds to begin accounting for the device stiffness in models that leverage the isometric and isotensional ensembles, respectively, depending on the regime of extension being considered.
Specifically, sufficiently large extensions can nonlinearly deform a stiff device, and sufficiently small extensions can cause nonlinear vibrations in a compliant device.
Correspondingly, future work should consider extending this asymptotic approach to account for the possible large deformation and nonlinear vibration of the device.

\begin{acknowledgments}
This work was supported by the Laboratory Directed Research and Development program at Sandia National Laboratories under Project No. 222398.
Sandia National Laboratories is a multi-mission laboratory managed and operated by National Technology and Engineering Solutions of Sandia, LLC., a wholly owned subsidiary of Honeywell International, Inc., for the U.S. Department of Energy's National Nuclear Security Administration under Contract No. DE-NA0003525.
Any subjective views or opinions expressed in the paper do not necessarily represent the views of the U.S. Department of Energy or the U.S. Government.
The U.S. Government retains and the publisher, by accepting the article for publication, acknowledges that the U.S. Government retains a nonexclusive, paid-up, irrevocable, world-wide license to publish or reproduce the published form of this manuscript, or allow others to do so, for U.S. Government purposes.
\end{acknowledgments}




\smallskip

\bibliography{main}

\begin{thebibliography}{71}%
\makeatletter
\providecommand \@ifxundefined [1]{%
 \@ifx{#1\undefined}
}%
\providecommand \@ifnum [1]{%
 \ifnum #1\expandafter \@firstoftwo
 \else \expandafter \@secondoftwo
 \fi
}%
\providecommand \@ifx [1]{%
 \ifx #1\expandafter \@firstoftwo
 \else \expandafter \@secondoftwo
 \fi
}%
\providecommand \natexlab [1]{#1}%
\providecommand \enquote  [1]{``#1''}%
\providecommand \bibnamefont  [1]{#1}%
\providecommand \bibfnamefont [1]{#1}%
\providecommand \citenamefont [1]{#1}%
\providecommand \href@noop [0]{\@secondoftwo}%
\providecommand \href [0]{\begingroup \@sanitize@url \@href}%
\providecommand \@href[1]{\@@startlink{#1}\@@href}%
\providecommand \@@href[1]{\endgroup#1\@@endlink}%
\providecommand \@sanitize@url [0]{\catcode `\\12\catcode `\$12\catcode
  `\&12\catcode `\#12\catcode `\^12\catcode `\_12\catcode `\%12\relax}%
\providecommand \@@startlink[1]{}%
\providecommand \@@endlink[0]{}%
\providecommand \url  [0]{\begingroup\@sanitize@url \@url }%
\providecommand \@url [1]{\endgroup\@href {#1}{\urlprefix }}%
\providecommand \urlprefix  [0]{URL }%
\providecommand \Eprint [0]{\href }%
\providecommand \doibase [0]{https://doi.org/}%
\providecommand \selectlanguage [0]{\@gobble}%
\providecommand \bibinfo  [0]{\@secondoftwo}%
\providecommand \bibfield  [0]{\@secondoftwo}%
\providecommand \translation [1]{[#1]}%
\providecommand \BibitemOpen [0]{}%
\providecommand \bibitemStop [0]{}%
\providecommand \bibitemNoStop [0]{.\EOS\space}%
\providecommand \EOS [0]{\spacefactor3000\relax}%
\providecommand \BibitemShut  [1]{\csname bibitem#1\endcsname}%
\let\auto@bib@innerbib\@empty
\bibitem [{\citenamefont {Binnig}\ \emph {et~al.}(1986)\citenamefont {Binnig},
  \citenamefont {Quate},\ and\ \citenamefont {Gerber}}]{binnig1986atomic}%
  \BibitemOpen
  \bibfield  {author} {\bibinfo {author} {\bibfnamefont {G.}~\bibnamefont
  {Binnig}}, \bibinfo {author} {\bibfnamefont {C.~F.}\ \bibnamefont {Quate}},\
  and\ \bibinfo {author} {\bibfnamefont {C.}~\bibnamefont {Gerber}},\
  }\bibfield  {title} {\bibinfo {title} {Atomic force microscope},\ }\href
  {https://doi.org/10.1103/PhysRevLett.56.930} {\bibfield  {journal} {\bibinfo
  {journal} {Physical Review Letters}\ }\textbf {\bibinfo {volume} {56}},\
  \bibinfo {pages} {930} (\bibinfo {year} {1986})}\BibitemShut {NoStop}%
\bibitem [{\citenamefont {Meyer}\ and\ \citenamefont
  {Amer}(1988)}]{meyer1988novel}%
  \BibitemOpen
  \bibfield  {author} {\bibinfo {author} {\bibfnamefont {G.}~\bibnamefont
  {Meyer}}\ and\ \bibinfo {author} {\bibfnamefont {N.~M.}\ \bibnamefont
  {Amer}},\ }\bibfield  {title} {\bibinfo {title} {Novel optical approach to
  atomic force microscopy},\ }\href {https://doi.org/10.1063/1.100061}
  {\bibfield  {journal} {\bibinfo  {journal} {Applied Physics Letters}\
  }\textbf {\bibinfo {volume} {53}},\ \bibinfo {pages} {1045} (\bibinfo {year}
  {1988})}\BibitemShut {NoStop}%
\bibitem [{\citenamefont {Florin}\ \emph {et~al.}(1994)\citenamefont {Florin},
  \citenamefont {Moy},\ and\ \citenamefont {Gaub}}]{florin1994adhesion}%
  \BibitemOpen
  \bibfield  {author} {\bibinfo {author} {\bibfnamefont {E.-L.}\ \bibnamefont
  {Florin}}, \bibinfo {author} {\bibfnamefont {V.~T.}\ \bibnamefont {Moy}},\
  and\ \bibinfo {author} {\bibfnamefont {H.~E.}\ \bibnamefont {Gaub}},\
  }\bibfield  {title} {\bibinfo {title} {Adhesion forces between individual
  ligand-receptor pairs},\ }\href {https://doi.org/10.1126/science.8153628}
  {\bibfield  {journal} {\bibinfo  {journal} {Science}\ }\textbf {\bibinfo
  {volume} {264}},\ \bibinfo {pages} {415} (\bibinfo {year}
  {1994})}\BibitemShut {NoStop}%
\bibitem [{\citenamefont {Lee}\ \emph {et~al.}(1994)\citenamefont {Lee},
  \citenamefont {Chrisey},\ and\ \citenamefont {Colton}}]{lee1994direct}%
  \BibitemOpen
  \bibfield  {author} {\bibinfo {author} {\bibfnamefont {G.~U.}\ \bibnamefont
  {Lee}}, \bibinfo {author} {\bibfnamefont {L.~A.}\ \bibnamefont {Chrisey}},\
  and\ \bibinfo {author} {\bibfnamefont {R.~J.}\ \bibnamefont {Colton}},\
  }\bibfield  {title} {\bibinfo {title} {Direct measurement of the forces
  between complementary strands of {DNA}},\ }\href
  {https://doi.org/10.1126/science.7973628} {\bibfield  {journal} {\bibinfo
  {journal} {Science}\ }\textbf {\bibinfo {volume} {266}},\ \bibinfo {pages}
  {771} (\bibinfo {year} {1994})}\BibitemShut {NoStop}%
\bibitem [{\citenamefont {Ortiz}\ and\ \citenamefont
  {Hadziioannou}(1999)}]{ortiz1999entropic}%
  \BibitemOpen
  \bibfield  {author} {\bibinfo {author} {\bibfnamefont {C.}~\bibnamefont
  {Ortiz}}\ and\ \bibinfo {author} {\bibfnamefont {G.}~\bibnamefont
  {Hadziioannou}},\ }\bibfield  {title} {\bibinfo {title} {Entropic elasticity
  of single polymer chains of poly (methacrylic acid) measured by atomic force
  microscopy},\ }\href {https://doi.org/10.1021/ma981245n} {\bibfield
  {journal} {\bibinfo  {journal} {Macromolecules}\ }\textbf {\bibinfo {volume}
  {32}},\ \bibinfo {pages} {780} (\bibinfo {year} {1999})}\BibitemShut
  {NoStop}%
\bibitem [{\citenamefont {Giessibl}(2003)}]{giessibl2003advances}%
  \BibitemOpen
  \bibfield  {author} {\bibinfo {author} {\bibfnamefont {F.~J.}\ \bibnamefont
  {Giessibl}},\ }\bibfield  {title} {\bibinfo {title} {Advances in atomic force
  microscopy},\ }\href {https://doi.org/10.1103/RevModPhys.75.949} {\bibfield
  {journal} {\bibinfo  {journal} {Reviews of Modern Physics}\ }\textbf
  {\bibinfo {volume} {75}},\ \bibinfo {pages} {949} (\bibinfo {year}
  {2003})}\BibitemShut {NoStop}%
\bibitem [{\citenamefont {Dudko}\ \emph {et~al.}(2008)\citenamefont {Dudko},
  \citenamefont {Hummer},\ and\ \citenamefont {Szabo}}]{dudko2008theory}%
  \BibitemOpen
  \bibfield  {author} {\bibinfo {author} {\bibfnamefont {O.~K.}\ \bibnamefont
  {Dudko}}, \bibinfo {author} {\bibfnamefont {G.}~\bibnamefont {Hummer}},\ and\
  \bibinfo {author} {\bibfnamefont {A.}~\bibnamefont {Szabo}},\ }\bibfield
  {title} {\bibinfo {title} {Theory, analysis, and interpretation of
  single-molecule force spectroscopy experiments},\ }\href
  {https://doi.org/10.1073/pnas.0806085105} {\bibfield  {journal} {\bibinfo
  {journal} {Proceedings of the National Academy of Sciences U.S.A.}\ }\textbf
  {\bibinfo {volume} {105}},\ \bibinfo {pages} {15755} (\bibinfo {year}
  {2008})}\BibitemShut {NoStop}%
\bibitem [{\citenamefont {Abkenar}\ \emph {et~al.}(2017)\citenamefont
  {Abkenar}, \citenamefont {Gray},\ and\ \citenamefont
  {Zaccone}}]{abkenar2017dissociation}%
  \BibitemOpen
  \bibfield  {author} {\bibinfo {author} {\bibfnamefont {M.}~\bibnamefont
  {Abkenar}}, \bibinfo {author} {\bibfnamefont {T.~H.}\ \bibnamefont {Gray}},\
  and\ \bibinfo {author} {\bibfnamefont {A.}~\bibnamefont {Zaccone}},\
  }\bibfield  {title} {\bibinfo {title} {Dissociation rates from
  single-molecule pulling experiments under large thermal fluctuations or large
  applied force},\ }\href {https://doi.org/10.1103/PhysRevE.95.042413}
  {\bibfield  {journal} {\bibinfo  {journal} {Physical Review E}\ }\textbf
  {\bibinfo {volume} {95}},\ \bibinfo {pages} {042413} (\bibinfo {year}
  {2017})}\BibitemShut {NoStop}%
\bibitem [{\citenamefont
  {Ashkin}(1970{\natexlab{a}})}]{ashkin1970acceleration}%
  \BibitemOpen
  \bibfield  {author} {\bibinfo {author} {\bibfnamefont {A.}~\bibnamefont
  {Ashkin}},\ }\bibfield  {title} {\bibinfo {title} {Acceleration and trapping
  of particles by radiation pressure},\ }\href
  {https://doi.org/10.1103/PhysRevLett.24.156} {\bibfield  {journal} {\bibinfo
  {journal} {Physical Review Letters}\ }\textbf {\bibinfo {volume} {24}},\
  \bibinfo {pages} {156} (\bibinfo {year} {1970}{\natexlab{a}})}\BibitemShut
  {NoStop}%
\bibitem [{\citenamefont {Ashkin}(1970{\natexlab{b}})}]{ashkin1970atomic}%
  \BibitemOpen
  \bibfield  {author} {\bibinfo {author} {\bibfnamefont {A.}~\bibnamefont
  {Ashkin}},\ }\bibfield  {title} {\bibinfo {title} {Atomic-beam deflection by
  resonance-radiation pressure},\ }\href
  {https://doi.org/10.1103/PhysRevLett.25.1321} {\bibfield  {journal} {\bibinfo
   {journal} {Physical Review Letters}\ }\textbf {\bibinfo {volume} {25}},\
  \bibinfo {pages} {1321} (\bibinfo {year} {1970}{\natexlab{b}})}\BibitemShut
  {NoStop}%
\bibitem [{\citenamefont {Chu}\ \emph {et~al.}(1985)\citenamefont {Chu},
  \citenamefont {Hollberg}, \citenamefont {Bjorkholm}, \citenamefont {Cable},\
  and\ \citenamefont {Ashkin}}]{chu1985three}%
  \BibitemOpen
  \bibfield  {author} {\bibinfo {author} {\bibfnamefont {S.}~\bibnamefont
  {Chu}}, \bibinfo {author} {\bibfnamefont {L.}~\bibnamefont {Hollberg}},
  \bibinfo {author} {\bibfnamefont {J.~E.}\ \bibnamefont {Bjorkholm}}, \bibinfo
  {author} {\bibfnamefont {A.}~\bibnamefont {Cable}},\ and\ \bibinfo {author}
  {\bibfnamefont {A.}~\bibnamefont {Ashkin}},\ }\bibfield  {title} {\bibinfo
  {title} {Three-dimensional viscous confinement and cooling of atoms by
  resonance radiation pressure},\ }\href
  {https://doi.org/10.1103/PhysRevLett.55.48} {\bibfield  {journal} {\bibinfo
  {journal} {Physical Review Letters}\ }\textbf {\bibinfo {volume} {55}},\
  \bibinfo {pages} {48} (\bibinfo {year} {1985})}\BibitemShut {NoStop}%
\bibitem [{\citenamefont {Chu}\ \emph {et~al.}(1986)\citenamefont {Chu},
  \citenamefont {Bjorkholm}, \citenamefont {Ashkin},\ and\ \citenamefont
  {Cable}}]{chu1986experimental}%
  \BibitemOpen
  \bibfield  {author} {\bibinfo {author} {\bibfnamefont {S.}~\bibnamefont
  {Chu}}, \bibinfo {author} {\bibfnamefont {J.}~\bibnamefont {Bjorkholm}},
  \bibinfo {author} {\bibfnamefont {A.}~\bibnamefont {Ashkin}},\ and\ \bibinfo
  {author} {\bibfnamefont {A.}~\bibnamefont {Cable}},\ }\bibfield  {title}
  {\bibinfo {title} {Experimental observation of optically trapped atoms},\
  }\href {https://doi.org/10.1103/PhysRevLett.57.314} {\bibfield  {journal}
  {\bibinfo  {journal} {Physical Review Letters}\ }\textbf {\bibinfo {volume}
  {57}},\ \bibinfo {pages} {314} (\bibinfo {year} {1986})}\BibitemShut
  {NoStop}%
\bibitem [{\citenamefont {Ashkin}\ and\ \citenamefont
  {Dziedzic}(1987)}]{ashkin1987optical}%
  \BibitemOpen
  \bibfield  {author} {\bibinfo {author} {\bibfnamefont {A.}~\bibnamefont
  {Ashkin}}\ and\ \bibinfo {author} {\bibfnamefont {J.~M.}\ \bibnamefont
  {Dziedzic}},\ }\bibfield  {title} {\bibinfo {title} {Optical trapping and
  manipulation of viruses and bacteria},\ }\href
  {https://doi.org/10.1126/science.3547653} {\bibfield  {journal} {\bibinfo
  {journal} {Science}\ }\textbf {\bibinfo {volume} {235}},\ \bibinfo {pages}
  {1517} (\bibinfo {year} {1987})}\BibitemShut {NoStop}%
\bibitem [{\citenamefont {Grier}(2003)}]{grier2003revolution}%
  \BibitemOpen
  \bibfield  {author} {\bibinfo {author} {\bibfnamefont {D.~G.}\ \bibnamefont
  {Grier}},\ }\bibfield  {title} {\bibinfo {title} {A revolution in optical
  manipulation},\ }\href {https://doi.org/10.1038/nature01935} {\bibfield
  {journal} {\bibinfo  {journal} {Nature}\ }\textbf {\bibinfo {volume} {424}},\
  \bibinfo {pages} {810} (\bibinfo {year} {2003})}\BibitemShut {NoStop}%
\bibitem [{\citenamefont {de~Lorenzo}\ \emph {et~al.}(2015)\citenamefont
  {de~Lorenzo}, \citenamefont {Ribezzi-Crivellari}, \citenamefont
  {Arias-Gonzalez}, \citenamefont {Smith},\ and\ \citenamefont
  {Ritort}}]{de2015temperature}%
  \BibitemOpen
  \bibfield  {author} {\bibinfo {author} {\bibfnamefont {S.}~\bibnamefont
  {de~Lorenzo}}, \bibinfo {author} {\bibfnamefont {M.}~\bibnamefont
  {Ribezzi-Crivellari}}, \bibinfo {author} {\bibfnamefont {J.~R.}\ \bibnamefont
  {Arias-Gonzalez}}, \bibinfo {author} {\bibfnamefont {S.~B.}\ \bibnamefont
  {Smith}},\ and\ \bibinfo {author} {\bibfnamefont {F.}~\bibnamefont
  {Ritort}},\ }\bibfield  {title} {\bibinfo {title} {A temperature-jump optical
  trap for single-molecule manipulation},\ }\href
  {https://doi.org/10.1016/j.bpj.2015.05.017} {\bibfield  {journal} {\bibinfo
  {journal} {Biophysical Journal}\ }\textbf {\bibinfo {volume} {108}},\
  \bibinfo {pages} {2854} (\bibinfo {year} {2015})}\BibitemShut {NoStop}%
\bibitem [{\citenamefont {Gieseler}\ \emph {et~al.}(2021)\citenamefont
  {Gieseler}, \citenamefont {Gomez-Solano}, \citenamefont {Magazz{\`u}},
  \citenamefont {Castillo}, \citenamefont {Garc{\'\i}a}, \citenamefont
  {Gironella-Torrent}, \citenamefont {Viader-Godoy}, \citenamefont {Ritort},
  \citenamefont {Pesce}, \citenamefont {Arzola} \emph
  {et~al.}}]{gieseler2021optical}%
  \BibitemOpen
  \bibfield  {author} {\bibinfo {author} {\bibfnamefont {J.}~\bibnamefont
  {Gieseler}}, \bibinfo {author} {\bibfnamefont {J.~R.}\ \bibnamefont
  {Gomez-Solano}}, \bibinfo {author} {\bibfnamefont {A.}~\bibnamefont
  {Magazz{\`u}}}, \bibinfo {author} {\bibfnamefont {I.~P.}\ \bibnamefont
  {Castillo}}, \bibinfo {author} {\bibfnamefont {L.~P.}\ \bibnamefont
  {Garc{\'\i}a}}, \bibinfo {author} {\bibfnamefont {M.}~\bibnamefont
  {Gironella-Torrent}}, \bibinfo {author} {\bibfnamefont {X.}~\bibnamefont
  {Viader-Godoy}}, \bibinfo {author} {\bibfnamefont {F.}~\bibnamefont
  {Ritort}}, \bibinfo {author} {\bibfnamefont {G.}~\bibnamefont {Pesce}},
  \bibinfo {author} {\bibfnamefont {A.~V.}\ \bibnamefont {Arzola}}, \emph
  {et~al.},\ }\bibfield  {title} {\bibinfo {title} {{Optical tweezers -- From
  calibration to applications: A tutorial}},\ }\href
  {https://doi.org/10.1364/AOP.394888} {\bibfield  {journal} {\bibinfo
  {journal} {Advances in Optics and Photonics}\ }\textbf {\bibinfo {volume}
  {13}},\ \bibinfo {pages} {74} (\bibinfo {year} {2021})}\BibitemShut {NoStop}%
\bibitem [{\citenamefont {Volpe}\ \emph {et~al.}(2023)\citenamefont {Volpe},
  \citenamefont {Marag{\`o}}, \citenamefont {Rubinsztein-Dunlop}, \citenamefont
  {Pesce}, \citenamefont {Stilgoe}, \citenamefont {Volpe}, \citenamefont
  {Tkachenko}, \citenamefont {Truong}, \citenamefont {Chormaic}, \citenamefont
  {Kalantarifard} \emph {et~al.}}]{volpe2023roadmap}%
  \BibitemOpen
  \bibfield  {author} {\bibinfo {author} {\bibfnamefont {G.}~\bibnamefont
  {Volpe}}, \bibinfo {author} {\bibfnamefont {O.~M.}\ \bibnamefont
  {Marag{\`o}}}, \bibinfo {author} {\bibfnamefont {H.}~\bibnamefont
  {Rubinsztein-Dunlop}}, \bibinfo {author} {\bibfnamefont {G.}~\bibnamefont
  {Pesce}}, \bibinfo {author} {\bibfnamefont {A.~B.}\ \bibnamefont {Stilgoe}},
  \bibinfo {author} {\bibfnamefont {G.}~\bibnamefont {Volpe}}, \bibinfo
  {author} {\bibfnamefont {G.}~\bibnamefont {Tkachenko}}, \bibinfo {author}
  {\bibfnamefont {V.~G.}\ \bibnamefont {Truong}}, \bibinfo {author}
  {\bibfnamefont {S.~N.}\ \bibnamefont {Chormaic}}, \bibinfo {author}
  {\bibfnamefont {F.}~\bibnamefont {Kalantarifard}}, \emph {et~al.},\
  }\bibfield  {title} {\bibinfo {title} {Roadmap for optical tweezers},\ }\href
  {https://doi.org/10.1088/2515-7647/acb57b} {\bibfield  {journal} {\bibinfo
  {journal} {Journal of Physics: Photonics}\ }\textbf {\bibinfo {volume} {5}},\
  \bibinfo {pages} {022501} (\bibinfo {year} {2023})}\BibitemShut {NoStop}%
\bibitem [{\citenamefont {Smith}\ \emph {et~al.}(1996)\citenamefont {Smith},
  \citenamefont {Cui},\ and\ \citenamefont
  {Bustamante}}]{smith1996overstretching}%
  \BibitemOpen
  \bibfield  {author} {\bibinfo {author} {\bibfnamefont {S.~B.}\ \bibnamefont
  {Smith}}, \bibinfo {author} {\bibfnamefont {Y.}~\bibnamefont {Cui}},\ and\
  \bibinfo {author} {\bibfnamefont {C.}~\bibnamefont {Bustamante}},\ }\bibfield
   {title} {\bibinfo {title} {{Overstretching B-DNA: The elastic response of
  individual double-stranded and single-stranded {DNA} molecules}},\ }\href
  {https://doi.org/10.1126/science.271.5250.795} {\bibfield  {journal}
  {\bibinfo  {journal} {Science}\ }\textbf {\bibinfo {volume} {271}},\ \bibinfo
  {pages} {795} (\bibinfo {year} {1996})}\BibitemShut {NoStop}%
\bibitem [{\citenamefont {Mehta}\ \emph {et~al.}(1999)\citenamefont {Mehta},
  \citenamefont {Rief}, \citenamefont {Spudich}, \citenamefont {Smith},\ and\
  \citenamefont {Simmons}}]{mehta1999single}%
  \BibitemOpen
  \bibfield  {author} {\bibinfo {author} {\bibfnamefont {A.~D.}\ \bibnamefont
  {Mehta}}, \bibinfo {author} {\bibfnamefont {M.}~\bibnamefont {Rief}},
  \bibinfo {author} {\bibfnamefont {J.~A.}\ \bibnamefont {Spudich}}, \bibinfo
  {author} {\bibfnamefont {D.~A.}\ \bibnamefont {Smith}},\ and\ \bibinfo
  {author} {\bibfnamefont {R.~M.}\ \bibnamefont {Simmons}},\ }\bibfield
  {title} {\bibinfo {title} {Single-molecule biomechanics with optical
  methods},\ }\href {https://doi.org/10.1126/science.283.5408.1689} {\bibfield
  {journal} {\bibinfo  {journal} {Science}\ }\textbf {\bibinfo {volume}
  {283}},\ \bibinfo {pages} {1689} (\bibinfo {year} {1999})}\BibitemShut
  {NoStop}%
\bibitem [{\citenamefont {Woodside}\ \emph {et~al.}(2006)\citenamefont
  {Woodside}, \citenamefont {Anthony}, \citenamefont {Behnke-Parks},
  \citenamefont {Larizadeh}, \citenamefont {Herschlag},\ and\ \citenamefont
  {Block}}]{woodside2006direct}%
  \BibitemOpen
  \bibfield  {author} {\bibinfo {author} {\bibfnamefont {M.~T.}\ \bibnamefont
  {Woodside}}, \bibinfo {author} {\bibfnamefont {P.~C.}\ \bibnamefont
  {Anthony}}, \bibinfo {author} {\bibfnamefont {W.~M.}\ \bibnamefont
  {Behnke-Parks}}, \bibinfo {author} {\bibfnamefont {K.}~\bibnamefont
  {Larizadeh}}, \bibinfo {author} {\bibfnamefont {D.}~\bibnamefont
  {Herschlag}},\ and\ \bibinfo {author} {\bibfnamefont {S.~M.}\ \bibnamefont
  {Block}},\ }\bibfield  {title} {\bibinfo {title} {Direct measurement of the
  full, sequence-dependent folding landscape of a nucleic acid},\ }\href
  {https://doi.org/10.1126/science.1133601} {\bibfield  {journal} {\bibinfo
  {journal} {Science}\ }\textbf {\bibinfo {volume} {314}},\ \bibinfo {pages}
  {1001} (\bibinfo {year} {2006})}\BibitemShut {NoStop}%
\bibitem [{\citenamefont {Woodside}\ and\ \citenamefont
  {Block}(2014)}]{woodside2014reconstructing}%
  \BibitemOpen
  \bibfield  {author} {\bibinfo {author} {\bibfnamefont {M.~T.}\ \bibnamefont
  {Woodside}}\ and\ \bibinfo {author} {\bibfnamefont {S.~M.}\ \bibnamefont
  {Block}},\ }\bibfield  {title} {\bibinfo {title} {Reconstructing folding
  energy landscapes by single-molecule force spectroscopy},\ }\href
  {https://doi.org/10.1146/annurev-biophys-051013-022754} {\bibfield  {journal}
  {\bibinfo  {journal} {Annual Review of Biophysics}\ }\textbf {\bibinfo
  {volume} {43}},\ \bibinfo {pages} {19} (\bibinfo {year} {2014})}\BibitemShut
  {NoStop}%
\bibitem [{\citenamefont {Dutta}\ \emph {et~al.}(2016)\citenamefont {Dutta},
  \citenamefont {Armitage},\ and\ \citenamefont
  {Lyubchenko}}]{dutta2016probing}%
  \BibitemOpen
  \bibfield  {author} {\bibinfo {author} {\bibfnamefont {S.}~\bibnamefont
  {Dutta}}, \bibinfo {author} {\bibfnamefont {B.~A.}\ \bibnamefont
  {Armitage}},\ and\ \bibinfo {author} {\bibfnamefont {Y.~L.}\ \bibnamefont
  {Lyubchenko}},\ }\bibfield  {title} {\bibinfo {title} {{Probing of
  miniPEG$\gamma$-PNA--DNA hybrid duplex stability with AFM force
  spectroscopy}},\ }\href {https://doi.org/10.1021/acs.biochem.5b01250}
  {\bibfield  {journal} {\bibinfo  {journal} {Biochemistry}\ }\textbf {\bibinfo
  {volume} {55}},\ \bibinfo {pages} {1523} (\bibinfo {year}
  {2016})}\BibitemShut {NoStop}%
\bibitem [{\citenamefont {Bustamante}\ \emph {et~al.}(2020)\citenamefont
  {Bustamante}, \citenamefont {Alexander}, \citenamefont {Maciuba},\ and\
  \citenamefont {Kaiser}}]{bustamante2020single}%
  \BibitemOpen
  \bibfield  {author} {\bibinfo {author} {\bibfnamefont {C.}~\bibnamefont
  {Bustamante}}, \bibinfo {author} {\bibfnamefont {L.}~\bibnamefont
  {Alexander}}, \bibinfo {author} {\bibfnamefont {K.}~\bibnamefont {Maciuba}},\
  and\ \bibinfo {author} {\bibfnamefont {C.~M.}\ \bibnamefont {Kaiser}},\
  }\bibfield  {title} {\bibinfo {title} {Single-molecule studies of protein
  folding with optical tweezers},\ }\href
  {https://doi.org/10.1146/annurev-biochem-013118-111442} {\bibfield  {journal}
  {\bibinfo  {journal} {Annual Review of Biochemistry}\ }\textbf {\bibinfo
  {volume} {89}},\ \bibinfo {pages} {443} (\bibinfo {year} {2020})}\BibitemShut
  {NoStop}%
\bibitem [{\citenamefont {Bustamante}\ \emph {et~al.}(2021)\citenamefont
  {Bustamante}, \citenamefont {Chemla}, \citenamefont {Liu},\ and\
  \citenamefont {Wang}}]{bustamante2021optical}%
  \BibitemOpen
  \bibfield  {author} {\bibinfo {author} {\bibfnamefont {C.~J.}\ \bibnamefont
  {Bustamante}}, \bibinfo {author} {\bibfnamefont {Y.~R.}\ \bibnamefont
  {Chemla}}, \bibinfo {author} {\bibfnamefont {S.}~\bibnamefont {Liu}},\ and\
  \bibinfo {author} {\bibfnamefont {M.~D.}\ \bibnamefont {Wang}},\ }\bibfield
  {title} {\bibinfo {title} {Optical tweezers in single-molecule biophysics},\
  }\href {https://doi.org/10.1038/s43586-021-00021-6} {\bibfield  {journal}
  {\bibinfo  {journal} {Nature Reviews Methods Primers}\ }\textbf {\bibinfo
  {volume} {1}},\ \bibinfo {pages} {25} (\bibinfo {year} {2021})}\BibitemShut
  {NoStop}%
\bibitem [{\citenamefont {Garcia-Manyes}\ and\ \citenamefont
  {Beedle}(2017)}]{garcia2017steering}%
  \BibitemOpen
  \bibfield  {author} {\bibinfo {author} {\bibfnamefont {S.}~\bibnamefont
  {Garcia-Manyes}}\ and\ \bibinfo {author} {\bibfnamefont {A.~E.}\ \bibnamefont
  {Beedle}},\ }\bibfield  {title} {\bibinfo {title} {Steering chemical
  reactions with force},\ }\href {https://doi.org/10.1038/s41570-017-0083}
  {\bibfield  {journal} {\bibinfo  {journal} {Nature Reviews Chemistry}\
  }\textbf {\bibinfo {volume} {1}},\ \bibinfo {pages} {0083} (\bibinfo {year}
  {2017})}\BibitemShut {NoStop}%
\bibitem [{\citenamefont {Do}\ and\ \citenamefont
  {Fri\v{s}\v{c}i\'{c}}(2017)}]{do2017mechanochemistry}%
  \BibitemOpen
  \bibfield  {author} {\bibinfo {author} {\bibfnamefont {J.-L.}\ \bibnamefont
  {Do}}\ and\ \bibinfo {author} {\bibfnamefont {T.}~\bibnamefont
  {Fri\v{s}\v{c}i\'{c}}},\ }\bibfield  {title} {\bibinfo {title}
  {{Mechanochemistry: A force of synthesis}},\ }\href
  {https://doi.org/10.1021/acscentsci.6b00277} {\bibfield  {journal} {\bibinfo
  {journal} {ACS Central Science}\ }\textbf {\bibinfo {volume} {3}},\ \bibinfo
  {pages} {13} (\bibinfo {year} {2017})}\BibitemShut {NoStop}%
\bibitem [{\citenamefont {O'Neill}\ and\ \citenamefont
  {Boulatov}(2021)}]{o2021many}%
  \BibitemOpen
  \bibfield  {author} {\bibinfo {author} {\bibfnamefont {R.~T.}\ \bibnamefont
  {O'Neill}}\ and\ \bibinfo {author} {\bibfnamefont {R.}~\bibnamefont
  {Boulatov}},\ }\bibfield  {title} {\bibinfo {title} {The many flavours of
  mechanochemistry and its plausible conceptual underpinnings},\ }\href
  {https://doi.org/10.1038/s41570-020-00249-y} {\bibfield  {journal} {\bibinfo
  {journal} {Nature Reviews Chemistry}\ }\textbf {\bibinfo {volume} {5}},\
  \bibinfo {pages} {148} (\bibinfo {year} {2021})}\BibitemShut {NoStop}%
\bibitem [{\citenamefont {Black}\ \emph {et~al.}(2011)\citenamefont {Black},
  \citenamefont {Lenhardt},\ and\ \citenamefont {Craig}}]{black2011molecular}%
  \BibitemOpen
  \bibfield  {author} {\bibinfo {author} {\bibfnamefont {A.~L.}\ \bibnamefont
  {Black}}, \bibinfo {author} {\bibfnamefont {J.~M.}\ \bibnamefont
  {Lenhardt}},\ and\ \bibinfo {author} {\bibfnamefont {S.~L.}\ \bibnamefont
  {Craig}},\ }\bibfield  {title} {\bibinfo {title} {From molecular
  mechanochemistry to stress-responsive materials},\ }\href
  {https://doi.org/10.1039/C0JM02636K} {\bibfield  {journal} {\bibinfo
  {journal} {Journal of Materials Chemistry}\ }\textbf {\bibinfo {volume}
  {21}},\ \bibinfo {pages} {1655} (\bibinfo {year} {2011})}\BibitemShut
  {NoStop}%
\bibitem [{\citenamefont {Brantley}\ \emph {et~al.}(2013)\citenamefont
  {Brantley}, \citenamefont {Wiggins},\ and\ \citenamefont
  {Bielawski}}]{brantley2013polymer}%
  \BibitemOpen
  \bibfield  {author} {\bibinfo {author} {\bibfnamefont {J.~N.}\ \bibnamefont
  {Brantley}}, \bibinfo {author} {\bibfnamefont {K.~M.}\ \bibnamefont
  {Wiggins}},\ and\ \bibinfo {author} {\bibfnamefont {C.~W.}\ \bibnamefont
  {Bielawski}},\ }\bibfield  {title} {\bibinfo {title} {{Polymer
  mechanochemistry: The design and study of mechanophores}},\ }\href
  {https://doi.org/10.1002/pi.4350} {\bibfield  {journal} {\bibinfo  {journal}
  {Polymer International}\ }\textbf {\bibinfo {volume} {62}},\ \bibinfo {pages}
  {2} (\bibinfo {year} {2013})}\BibitemShut {NoStop}%
\bibitem [{\citenamefont {Vidavsky}\ \emph {et~al.}(2019)\citenamefont
  {Vidavsky}, \citenamefont {Yang}, \citenamefont {Abel}, \citenamefont
  {Agami}, \citenamefont {Diesendruck}, \citenamefont {Coates},\ and\
  \citenamefont {Silberstein}}]{vidavsky2019enabling}%
  \BibitemOpen
  \bibfield  {author} {\bibinfo {author} {\bibfnamefont {Y.}~\bibnamefont
  {Vidavsky}}, \bibinfo {author} {\bibfnamefont {S.~J.}\ \bibnamefont {Yang}},
  \bibinfo {author} {\bibfnamefont {B.~A.}\ \bibnamefont {Abel}}, \bibinfo
  {author} {\bibfnamefont {I.}~\bibnamefont {Agami}}, \bibinfo {author}
  {\bibfnamefont {C.~E.}\ \bibnamefont {Diesendruck}}, \bibinfo {author}
  {\bibfnamefont {G.~W.}\ \bibnamefont {Coates}},\ and\ \bibinfo {author}
  {\bibfnamefont {M.~N.}\ \bibnamefont {Silberstein}},\ }\bibfield  {title}
  {\bibinfo {title} {{Enabling room-temperature mechanochromic activation in a
  glassy polymer: Synthesis and characterization of spiropyran
  polycarbonate}},\ }\href {https://doi.org/10.1021/jacs.9b04229} {\bibfield
  {journal} {\bibinfo  {journal} {Journal of the American Chemical Society}\
  }\textbf {\bibinfo {volume} {141}},\ \bibinfo {pages} {10060} (\bibinfo
  {year} {2019})}\BibitemShut {NoStop}%
\bibitem [{\citenamefont {Chen}\ \emph {et~al.}(2021)\citenamefont {Chen},
  \citenamefont {Mellot}, \citenamefont {van Luijk}, \citenamefont {Creton},\
  and\ \citenamefont {Sijbesma}}]{chen2021mechanochemical}%
  \BibitemOpen
  \bibfield  {author} {\bibinfo {author} {\bibfnamefont {Y.}~\bibnamefont
  {Chen}}, \bibinfo {author} {\bibfnamefont {G.}~\bibnamefont {Mellot}},
  \bibinfo {author} {\bibfnamefont {D.}~\bibnamefont {van Luijk}}, \bibinfo
  {author} {\bibfnamefont {C.}~\bibnamefont {Creton}},\ and\ \bibinfo {author}
  {\bibfnamefont {R.~P.}\ \bibnamefont {Sijbesma}},\ }\bibfield  {title}
  {\bibinfo {title} {Mechanochemical tools for polymer materials},\ }\href
  {https://doi.org/10.1039/D0CS00940G} {\bibfield  {journal} {\bibinfo
  {journal} {Chemical Society Reviews}\ }\textbf {\bibinfo {volume} {50}},\
  \bibinfo {pages} {4100} (\bibinfo {year} {2021})}\BibitemShut {NoStop}%
\bibitem [{\citenamefont {Jayathilaka}\ \emph {et~al.}(2021)\citenamefont
  {Jayathilaka}, \citenamefont {Molley}, \citenamefont {Huang}, \citenamefont
  {Islam}, \citenamefont {Buche}, \citenamefont {Silberstein}, \citenamefont
  {Kruzic},\ and\ \citenamefont {Kilian}}]{jayathilaka2021force}%
  \BibitemOpen
  \bibfield  {author} {\bibinfo {author} {\bibfnamefont {P.~B.}\ \bibnamefont
  {Jayathilaka}}, \bibinfo {author} {\bibfnamefont {T.~G.}\ \bibnamefont
  {Molley}}, \bibinfo {author} {\bibfnamefont {Y.}~\bibnamefont {Huang}},
  \bibinfo {author} {\bibfnamefont {M.~S.}\ \bibnamefont {Islam}}, \bibinfo
  {author} {\bibfnamefont {M.~R.}\ \bibnamefont {Buche}}, \bibinfo {author}
  {\bibfnamefont {M.~N.}\ \bibnamefont {Silberstein}}, \bibinfo {author}
  {\bibfnamefont {J.~J.}\ \bibnamefont {Kruzic}},\ and\ \bibinfo {author}
  {\bibfnamefont {K.~A.}\ \bibnamefont {Kilian}},\ }\bibfield  {title}
  {\bibinfo {title} {Force-mediated molecule release from double network
  hydrogels},\ }\href {https://doi.org/10.1039/D1CC02726C} {\bibfield
  {journal} {\bibinfo  {journal} {Chemical Communications}\ }\textbf {\bibinfo
  {volume} {57}},\ \bibinfo {pages} {8484} (\bibinfo {year}
  {2021})}\BibitemShut {NoStop}%
\bibitem [{\citenamefont {Ghanem}\ \emph {et~al.}(2021)\citenamefont {Ghanem},
  \citenamefont {Basu}, \citenamefont {Behrou}, \citenamefont {Boechler},
  \citenamefont {Boydston}, \citenamefont {Craig}, \citenamefont {Lin},
  \citenamefont {Lynde}, \citenamefont {Nelson}, \citenamefont {Shen} \emph
  {et~al.}}]{ghanem2021role}%
  \BibitemOpen
  \bibfield  {author} {\bibinfo {author} {\bibfnamefont {M.~A.}\ \bibnamefont
  {Ghanem}}, \bibinfo {author} {\bibfnamefont {A.}~\bibnamefont {Basu}},
  \bibinfo {author} {\bibfnamefont {R.}~\bibnamefont {Behrou}}, \bibinfo
  {author} {\bibfnamefont {N.}~\bibnamefont {Boechler}}, \bibinfo {author}
  {\bibfnamefont {A.~J.}\ \bibnamefont {Boydston}}, \bibinfo {author}
  {\bibfnamefont {S.~L.}\ \bibnamefont {Craig}}, \bibinfo {author}
  {\bibfnamefont {Y.}~\bibnamefont {Lin}}, \bibinfo {author} {\bibfnamefont
  {B.~E.}\ \bibnamefont {Lynde}}, \bibinfo {author} {\bibfnamefont
  {A.}~\bibnamefont {Nelson}}, \bibinfo {author} {\bibfnamefont
  {H.}~\bibnamefont {Shen}}, \emph {et~al.},\ }\bibfield  {title} {\bibinfo
  {title} {The role of polymer mechanochemistry in responsive materials and
  additive manufacturing},\ }\href {https://doi.org/10.1038/s41578-020-00249-w}
  {\bibfield  {journal} {\bibinfo  {journal} {Nature Reviews Materials}\
  }\textbf {\bibinfo {volume} {6}},\ \bibinfo {pages} {84} (\bibinfo {year}
  {2021})}\BibitemShut {NoStop}%
\bibitem [{\citenamefont {Evans}(2001)}]{evans2001probing}%
  \BibitemOpen
  \bibfield  {author} {\bibinfo {author} {\bibfnamefont {E.}~\bibnamefont
  {Evans}},\ }\bibfield  {title} {\bibinfo {title} {Probing the relation
  between force-lifetime-and chemistry in single molecular bonds},\ }\href
  {https://doi.org/10.1146/annurev.biophys.30.1.105} {\bibfield  {journal}
  {\bibinfo  {journal} {Annual Review of Biophysics and Biomolecular
  Structure}\ }\textbf {\bibinfo {volume} {30}},\ \bibinfo {pages} {105}
  (\bibinfo {year} {2001})}\BibitemShut {NoStop}%
\bibitem [{\citenamefont {Liang}\ and\ \citenamefont
  {Fern{\'a}ndez}(2009)}]{liang2009mechanochemistry}%
  \BibitemOpen
  \bibfield  {author} {\bibinfo {author} {\bibfnamefont {J.}~\bibnamefont
  {Liang}}\ and\ \bibinfo {author} {\bibfnamefont {J.~M.}\ \bibnamefont
  {Fern{\'a}ndez}},\ }\bibfield  {title} {\bibinfo {title} {{Mechanochemistry:
  One bond at a time}},\ }\href {https://doi.org/10.1021/nn900294n} {\bibfield
  {journal} {\bibinfo  {journal} {ACS Nano}\ }\textbf {\bibinfo {volume} {3}},\
  \bibinfo {pages} {1628} (\bibinfo {year} {2009})}\BibitemShut {NoStop}%
\bibitem [{\citenamefont {Wang}\ \emph {et~al.}(2015)\citenamefont {Wang},
  \citenamefont {Kouznetsova}, \citenamefont {Niu}, \citenamefont {Ong},
  \citenamefont {Klukovich}, \citenamefont {Rheingold}, \citenamefont
  {Martinez},\ and\ \citenamefont {Craig}}]{wang2015inducing}%
  \BibitemOpen
  \bibfield  {author} {\bibinfo {author} {\bibfnamefont {J.}~\bibnamefont
  {Wang}}, \bibinfo {author} {\bibfnamefont {T.~B.}\ \bibnamefont
  {Kouznetsova}}, \bibinfo {author} {\bibfnamefont {Z.}~\bibnamefont {Niu}},
  \bibinfo {author} {\bibfnamefont {M.~T.}\ \bibnamefont {Ong}}, \bibinfo
  {author} {\bibfnamefont {H.~M.}\ \bibnamefont {Klukovich}}, \bibinfo {author}
  {\bibfnamefont {A.~L.}\ \bibnamefont {Rheingold}}, \bibinfo {author}
  {\bibfnamefont {T.~J.}\ \bibnamefont {Martinez}},\ and\ \bibinfo {author}
  {\bibfnamefont {S.~L.}\ \bibnamefont {Craig}},\ }\bibfield  {title} {\bibinfo
  {title} {Inducing and quantifying forbidden reactivity with single-molecule
  polymer mechanochemistry},\ }\href {https://doi.org/10.1038/nchem.2185}
  {\bibfield  {journal} {\bibinfo  {journal} {Nature Chemistry}\ }\textbf
  {\bibinfo {volume} {7}},\ \bibinfo {pages} {323} (\bibinfo {year}
  {2015})}\BibitemShut {NoStop}%
\bibitem [{\citenamefont {Gossweiler}\ \emph {et~al.}(2015)\citenamefont
  {Gossweiler}, \citenamefont {Kouznetsova},\ and\ \citenamefont
  {Craig}}]{gossweiler2015force}%
  \BibitemOpen
  \bibfield  {author} {\bibinfo {author} {\bibfnamefont {G.~R.}\ \bibnamefont
  {Gossweiler}}, \bibinfo {author} {\bibfnamefont {T.~B.}\ \bibnamefont
  {Kouznetsova}},\ and\ \bibinfo {author} {\bibfnamefont {S.~L.}\ \bibnamefont
  {Craig}},\ }\bibfield  {title} {\bibinfo {title} {Force-rate characterization
  of two spiropyran-based molecular force probes},\ }\href
  {https://doi.org/10.1021/jacs.5b02492} {\bibfield  {journal} {\bibinfo
  {journal} {Journal of the American Chemical Society}\ }\textbf {\bibinfo
  {volume} {137}},\ \bibinfo {pages} {6148} (\bibinfo {year}
  {2015})}\BibitemShut {NoStop}%
\bibitem [{\citenamefont {Zhang}\ \emph {et~al.}(2017)\citenamefont {Zhang},
  \citenamefont {Li}, \citenamefont {Lin}, \citenamefont {Gao}, \citenamefont
  {Tang}, \citenamefont {Su}, \citenamefont {Zhang}, \citenamefont {Xu},
  \citenamefont {Weng},\ and\ \citenamefont {Boulatov}}]{zhang2017multi}%
  \BibitemOpen
  \bibfield  {author} {\bibinfo {author} {\bibfnamefont {H.}~\bibnamefont
  {Zhang}}, \bibinfo {author} {\bibfnamefont {X.}~\bibnamefont {Li}}, \bibinfo
  {author} {\bibfnamefont {Y.}~\bibnamefont {Lin}}, \bibinfo {author}
  {\bibfnamefont {F.}~\bibnamefont {Gao}}, \bibinfo {author} {\bibfnamefont
  {Z.}~\bibnamefont {Tang}}, \bibinfo {author} {\bibfnamefont {P.}~\bibnamefont
  {Su}}, \bibinfo {author} {\bibfnamefont {W.}~\bibnamefont {Zhang}}, \bibinfo
  {author} {\bibfnamefont {Y.}~\bibnamefont {Xu}}, \bibinfo {author}
  {\bibfnamefont {W.}~\bibnamefont {Weng}},\ and\ \bibinfo {author}
  {\bibfnamefont {R.}~\bibnamefont {Boulatov}},\ }\bibfield  {title} {\bibinfo
  {title} {Multi-modal mechanophores based on cinnamate dimers},\ }\href
  {https://doi.org/10.1038/s41467-017-01412-8} {\bibfield  {journal} {\bibinfo
  {journal} {Nature Communications}\ }\textbf {\bibinfo {volume} {8}},\
  \bibinfo {pages} {1147} (\bibinfo {year} {2017})}\BibitemShut {NoStop}%
\bibitem [{\citenamefont {Akbulatov}\ and\ \citenamefont
  {Boulatov}(2017)}]{akbulatov2017experimental}%
  \BibitemOpen
  \bibfield  {author} {\bibinfo {author} {\bibfnamefont {S.}~\bibnamefont
  {Akbulatov}}\ and\ \bibinfo {author} {\bibfnamefont {R.}~\bibnamefont
  {Boulatov}},\ }\bibfield  {title} {\bibinfo {title} {Experimental polymer
  mechanochemistry and its interpretational frameworks},\ }\href
  {https://doi.org/10.1002/cphc.201601354} {\bibfield  {journal} {\bibinfo
  {journal} {ChemPhysChem}\ }\textbf {\bibinfo {volume} {18}},\ \bibinfo
  {pages} {1422} (\bibinfo {year} {2017})}\BibitemShut {NoStop}%
\bibitem [{\citenamefont {Barbee}\ \emph {et~al.}(2018)\citenamefont {Barbee},
  \citenamefont {Kouznetsova}, \citenamefont {Barrett}, \citenamefont
  {Gossweiler}, \citenamefont {Lin}, \citenamefont {Rastogi}, \citenamefont
  {Brittain},\ and\ \citenamefont {Craig}}]{barbee2018substituent}%
  \BibitemOpen
  \bibfield  {author} {\bibinfo {author} {\bibfnamefont {M.~H.}\ \bibnamefont
  {Barbee}}, \bibinfo {author} {\bibfnamefont {T.}~\bibnamefont {Kouznetsova}},
  \bibinfo {author} {\bibfnamefont {S.~L.}\ \bibnamefont {Barrett}}, \bibinfo
  {author} {\bibfnamefont {G.~R.}\ \bibnamefont {Gossweiler}}, \bibinfo
  {author} {\bibfnamefont {Y.}~\bibnamefont {Lin}}, \bibinfo {author}
  {\bibfnamefont {S.~K.}\ \bibnamefont {Rastogi}}, \bibinfo {author}
  {\bibfnamefont {W.~J.}\ \bibnamefont {Brittain}},\ and\ \bibinfo {author}
  {\bibfnamefont {S.~L.}\ \bibnamefont {Craig}},\ }\bibfield  {title} {\bibinfo
  {title} {Substituent effects and mechanism in a mechanochemical reaction},\
  }\href {https://doi.org/10.1021/jacs.8b09263} {\bibfield  {journal} {\bibinfo
   {journal} {Journal of the American Chemical Society}\ }\textbf {\bibinfo
  {volume} {140}},\ \bibinfo {pages} {12746} (\bibinfo {year}
  {2018})}\BibitemShut {NoStop}%
\bibitem [{\citenamefont {Sulkanen}\ \emph {et~al.}(2019)\citenamefont
  {Sulkanen}, \citenamefont {Sung}, \citenamefont {Robb}, \citenamefont
  {Moore}, \citenamefont {Sottos},\ and\ \citenamefont
  {Liu}}]{sulkanen2019spatially}%
  \BibitemOpen
  \bibfield  {author} {\bibinfo {author} {\bibfnamefont {A.~R.}\ \bibnamefont
  {Sulkanen}}, \bibinfo {author} {\bibfnamefont {J.}~\bibnamefont {Sung}},
  \bibinfo {author} {\bibfnamefont {M.~J.}\ \bibnamefont {Robb}}, \bibinfo
  {author} {\bibfnamefont {J.~S.}\ \bibnamefont {Moore}}, \bibinfo {author}
  {\bibfnamefont {N.~R.}\ \bibnamefont {Sottos}},\ and\ \bibinfo {author}
  {\bibfnamefont {G.-y.}\ \bibnamefont {Liu}},\ }\bibfield  {title} {\bibinfo
  {title} {Spatially selective and density-controlled activation of interfacial
  mechanophores},\ }\href {https://doi.org/10.1021/jacs.8b10257} {\bibfield
  {journal} {\bibinfo  {journal} {Journal of the American Chemical Society}\
  }\textbf {\bibinfo {volume} {141}},\ \bibinfo {pages} {4080} (\bibinfo {year}
  {2019})}\BibitemShut {NoStop}%
\bibitem [{\citenamefont {Liu}\ and\ \citenamefont
  {Vancso}(2020)}]{liu2020polymer}%
  \BibitemOpen
  \bibfield  {author} {\bibinfo {author} {\bibfnamefont {Y.}~\bibnamefont
  {Liu}}\ and\ \bibinfo {author} {\bibfnamefont {G.~J.}\ \bibnamefont
  {Vancso}},\ }\bibfield  {title} {\bibinfo {title} {{Polymer single chain
  imaging, molecular forces, and nanoscale processes by Atomic Force
  Microscopy: The ultimate proof of the macromolecular hypothesis}},\ }\href
  {https://doi.org/10.1016/j.progpolymsci.2020.101232} {\bibfield  {journal}
  {\bibinfo  {journal} {Progress in Polymer Science}\ }\textbf {\bibinfo
  {volume} {104}},\ \bibinfo {pages} {101232} (\bibinfo {year}
  {2020})}\BibitemShut {NoStop}%
\bibitem [{\citenamefont {Stratigaki}\ and\ \citenamefont
  {G{\"o}stl}(2020)}]{stratigaki2020methods}%
  \BibitemOpen
  \bibfield  {author} {\bibinfo {author} {\bibfnamefont {M.}~\bibnamefont
  {Stratigaki}}\ and\ \bibinfo {author} {\bibfnamefont {R.}~\bibnamefont
  {G{\"o}stl}},\ }\bibfield  {title} {\bibinfo {title} {Methods for exerting
  and sensing force in polymer materials using mechanophores},\ }\href
  {https://doi.org/10.1002/cplu.201900737} {\bibfield  {journal} {\bibinfo
  {journal} {ChemPlusChem}\ }\textbf {\bibinfo {volume} {85}},\ \bibinfo
  {pages} {1095} (\bibinfo {year} {2020})}\BibitemShut {NoStop}%
\bibitem [{\citenamefont {Moses}\ \emph {et~al.}(2004)\citenamefont {Moses},
  \citenamefont {Brewer}, \citenamefont {Lowe}, \citenamefont {Lappi},
  \citenamefont {Gilvey}, \citenamefont {Sauthier}, \citenamefont {Tenent},
  \citenamefont {Feldheim},\ and\ \citenamefont
  {Franzen}}]{moses2004characterization}%
  \BibitemOpen
  \bibfield  {author} {\bibinfo {author} {\bibfnamefont {S.}~\bibnamefont
  {Moses}}, \bibinfo {author} {\bibfnamefont {S.~H.}\ \bibnamefont {Brewer}},
  \bibinfo {author} {\bibfnamefont {L.~B.}\ \bibnamefont {Lowe}}, \bibinfo
  {author} {\bibfnamefont {S.~E.}\ \bibnamefont {Lappi}}, \bibinfo {author}
  {\bibfnamefont {L.~B.}\ \bibnamefont {Gilvey}}, \bibinfo {author}
  {\bibfnamefont {M.}~\bibnamefont {Sauthier}}, \bibinfo {author}
  {\bibfnamefont {R.~C.}\ \bibnamefont {Tenent}}, \bibinfo {author}
  {\bibfnamefont {D.~L.}\ \bibnamefont {Feldheim}},\ and\ \bibinfo {author}
  {\bibfnamefont {S.}~\bibnamefont {Franzen}},\ }\bibfield  {title} {\bibinfo
  {title} {Characterization of single-and double-stranded {DNA} on gold
  surfaces},\ }\href {https://doi.org/10.1021/la0492815} {\bibfield  {journal}
  {\bibinfo  {journal} {Langmuir}\ }\textbf {\bibinfo {volume} {20}},\ \bibinfo
  {pages} {11134} (\bibinfo {year} {2004})}\BibitemShut {NoStop}%
\bibitem [{\citenamefont {Huang}\ \emph {et~al.}(2017)\citenamefont {Huang},
  \citenamefont {Zhu}, \citenamefont {Wen}, \citenamefont {Wang}, \citenamefont
  {Qin}, \citenamefont {Cao}, \citenamefont {Ma},\ and\ \citenamefont
  {Wang}}]{huang2017single}%
  \BibitemOpen
  \bibfield  {author} {\bibinfo {author} {\bibfnamefont {W.}~\bibnamefont
  {Huang}}, \bibinfo {author} {\bibfnamefont {Z.}~\bibnamefont {Zhu}}, \bibinfo
  {author} {\bibfnamefont {J.}~\bibnamefont {Wen}}, \bibinfo {author}
  {\bibfnamefont {X.}~\bibnamefont {Wang}}, \bibinfo {author} {\bibfnamefont
  {M.}~\bibnamefont {Qin}}, \bibinfo {author} {\bibfnamefont {Y.}~\bibnamefont
  {Cao}}, \bibinfo {author} {\bibfnamefont {H.}~\bibnamefont {Ma}},\ and\
  \bibinfo {author} {\bibfnamefont {W.}~\bibnamefont {Wang}},\ }\bibfield
  {title} {\bibinfo {title} {Single molecule study of force-induced rotation of
  carbon--carbon double bonds in polymers},\ }\href
  {https://doi.org/10.1021/acsnano.6b07119} {\bibfield  {journal} {\bibinfo
  {journal} {ACS Nano}\ }\textbf {\bibinfo {volume} {11}},\ \bibinfo {pages}
  {194} (\bibinfo {year} {2017})}\BibitemShut {NoStop}%
\bibitem [{\citenamefont {Vidavsky}\ \emph {et~al.}(2020)\citenamefont
  {Vidavsky}, \citenamefont {Buche}, \citenamefont {Sparrow}, \citenamefont
  {Zhang}, \citenamefont {Yang}, \citenamefont {DiStasio},\ and\ \citenamefont
  {Silberstein}}]{vidavsky2020tuning}%
  \BibitemOpen
  \bibfield  {author} {\bibinfo {author} {\bibfnamefont {Y.}~\bibnamefont
  {Vidavsky}}, \bibinfo {author} {\bibfnamefont {M.~R.}\ \bibnamefont {Buche}},
  \bibinfo {author} {\bibfnamefont {Z.~M.}\ \bibnamefont {Sparrow}}, \bibinfo
  {author} {\bibfnamefont {X.}~\bibnamefont {Zhang}}, \bibinfo {author}
  {\bibfnamefont {S.~J.}\ \bibnamefont {Yang}}, \bibinfo {author}
  {\bibfnamefont {R.~A.}\ \bibnamefont {DiStasio}},\ and\ \bibinfo {author}
  {\bibfnamefont {M.~N.}\ \bibnamefont {Silberstein}},\ }\bibfield  {title}
  {\bibinfo {title} {{Tuning the mechanical properties of metallopolymers via
  ligand interactions: A combined experimental and theoretical study}},\ }\href
  {https://doi.org/10.1021/acs.macromol.9b02756} {\bibfield  {journal}
  {\bibinfo  {journal} {Macromolecules}\ }\textbf {\bibinfo {volume} {53}},\
  \bibinfo {pages} {2021} (\bibinfo {year} {2020})}\BibitemShut {NoStop}%
\bibitem [{\citenamefont {Hsissou}\ \emph {et~al.}(2021)\citenamefont
  {Hsissou}, \citenamefont {Abbout}, \citenamefont {Safi}, \citenamefont
  {Benhiba}, \citenamefont {Wazzan}, \citenamefont {Guo}, \citenamefont
  {Nouneh}, \citenamefont {Briche}, \citenamefont {Erramli}, \citenamefont
  {Touhami} \emph {et~al.}}]{hsissou2021synthesis}%
  \BibitemOpen
  \bibfield  {author} {\bibinfo {author} {\bibfnamefont {R.}~\bibnamefont
  {Hsissou}}, \bibinfo {author} {\bibfnamefont {S.}~\bibnamefont {Abbout}},
  \bibinfo {author} {\bibfnamefont {Z.}~\bibnamefont {Safi}}, \bibinfo {author}
  {\bibfnamefont {F.}~\bibnamefont {Benhiba}}, \bibinfo {author} {\bibfnamefont
  {N.}~\bibnamefont {Wazzan}}, \bibinfo {author} {\bibfnamefont
  {L.}~\bibnamefont {Guo}}, \bibinfo {author} {\bibfnamefont {K.}~\bibnamefont
  {Nouneh}}, \bibinfo {author} {\bibfnamefont {S.}~\bibnamefont {Briche}},
  \bibinfo {author} {\bibfnamefont {H.}~\bibnamefont {Erramli}}, \bibinfo
  {author} {\bibfnamefont {M.~E.}\ \bibnamefont {Touhami}}, \emph {et~al.},\
  }\bibfield  {title} {\bibinfo {title} {{Synthesis and anticorrosive
  properties of epoxy polymer for CS in [1 M] HCl solution: Electrochemical,
  AFM, DFT and MD simulations}},\ }\href
  {https://doi.org/10.1016/j.conbuildmat.2020.121454} {\bibfield  {journal}
  {\bibinfo  {journal} {Construction and Building Materials}\ }\textbf
  {\bibinfo {volume} {270}},\ \bibinfo {pages} {121454} (\bibinfo {year}
  {2021})}\BibitemShut {NoStop}%
\bibitem [{\citenamefont {Neumann}(2003)}]{neumann2003precise}%
  \BibitemOpen
  \bibfield  {author} {\bibinfo {author} {\bibfnamefont {R.~M.}\ \bibnamefont
  {Neumann}},\ }\bibfield  {title} {\bibinfo {title} {On the precise meaning of
  extension in the interpretation of polymer-chain stretching experiments},\
  }\href {https://doi.org/10.1016/S0006-3495(03)74760-2} {\bibfield  {journal}
  {\bibinfo  {journal} {Biophysical journal}\ }\textbf {\bibinfo {volume}
  {85}},\ \bibinfo {pages} {3418} (\bibinfo {year} {2003})}\BibitemShut
  {NoStop}%
\bibitem [{\citenamefont {Sinha}\ and\ \citenamefont
  {Samuel}(2005)}]{sinha2005inequivalence}%
  \BibitemOpen
  \bibfield  {author} {\bibinfo {author} {\bibfnamefont {S.}~\bibnamefont
  {Sinha}}\ and\ \bibinfo {author} {\bibfnamefont {J.}~\bibnamefont {Samuel}},\
  }\bibfield  {title} {\bibinfo {title} {Inequivalence of statistical ensembles
  in single molecule measurements},\ }\href
  {https://doi.org/10.1103/PhysRevE.71.021104} {\bibfield  {journal} {\bibinfo
  {journal} {Physical Review E}\ }\textbf {\bibinfo {volume} {71}},\ \bibinfo
  {pages} {021104} (\bibinfo {year} {2005})}\BibitemShut {NoStop}%
\bibitem [{\citenamefont {S{\"u}zen}\ \emph {et~al.}(2009)\citenamefont
  {S{\"u}zen}, \citenamefont {Sega},\ and\ \citenamefont
  {Holm}}]{suzen2009ensemble}%
  \BibitemOpen
  \bibfield  {author} {\bibinfo {author} {\bibfnamefont {M.}~\bibnamefont
  {S{\"u}zen}}, \bibinfo {author} {\bibfnamefont {M.}~\bibnamefont {Sega}},\
  and\ \bibinfo {author} {\bibfnamefont {C.}~\bibnamefont {Holm}},\ }\bibfield
  {title} {\bibinfo {title} {Ensemble inequivalence in single-molecule
  experiments},\ }\href {https://doi.org/10.1103/PhysRevE.79.051118} {\bibfield
   {journal} {\bibinfo  {journal} {Physical Review E}\ }\textbf {\bibinfo
  {volume} {79}},\ \bibinfo {pages} {051118} (\bibinfo {year}
  {2009})}\BibitemShut {NoStop}%
\bibitem [{\citenamefont {Manca}\ \emph {et~al.}(2014)\citenamefont {Manca},
  \citenamefont {Giordano}, \citenamefont {Palla},\ and\ \citenamefont
  {Cleri}}]{manca2014equivalence}%
  \BibitemOpen
  \bibfield  {author} {\bibinfo {author} {\bibfnamefont {F.}~\bibnamefont
  {Manca}}, \bibinfo {author} {\bibfnamefont {S.}~\bibnamefont {Giordano}},
  \bibinfo {author} {\bibfnamefont {P.~L.}\ \bibnamefont {Palla}},\ and\
  \bibinfo {author} {\bibfnamefont {F.}~\bibnamefont {Cleri}},\ }\bibfield
  {title} {\bibinfo {title} {On the equivalence of thermodynamics ensembles for
  flexible polymer chains},\ }\href
  {https://doi.org/10.1016/j.physa.2013.10.042} {\bibfield  {journal} {\bibinfo
   {journal} {Physica A: Statistical Mechanics and its Applications}\ }\textbf
  {\bibinfo {volume} {395}},\ \bibinfo {pages} {154} (\bibinfo {year}
  {2014})}\BibitemShut {NoStop}%
\bibitem [{\citenamefont {Dutta}\ and\ \citenamefont
  {Benetatos}(2018)}]{dutta2018inequivalence}%
  \BibitemOpen
  \bibfield  {author} {\bibinfo {author} {\bibfnamefont {S.}~\bibnamefont
  {Dutta}}\ and\ \bibinfo {author} {\bibfnamefont {P.}~\bibnamefont
  {Benetatos}},\ }\bibfield  {title} {\bibinfo {title} {Inequivalence of
  fixed-force and fixed-extension statistical ensembles for a flexible polymer
  tethered to a planar substrate},\ }\href {https://doi.org/10.1039/C8SM01321G}
  {\bibfield  {journal} {\bibinfo  {journal} {Soft Matter}\ }\textbf {\bibinfo
  {volume} {14}},\ \bibinfo {pages} {6857} (\bibinfo {year}
  {2018})}\BibitemShut {NoStop}%
\bibitem [{\citenamefont {Benedito}\ and\ \citenamefont
  {Giordano}(2018)}]{benedito2018isotensional}%
  \BibitemOpen
  \bibfield  {author} {\bibinfo {author} {\bibfnamefont {M.}~\bibnamefont
  {Benedito}}\ and\ \bibinfo {author} {\bibfnamefont {S.}~\bibnamefont
  {Giordano}},\ }\bibfield  {title} {\bibinfo {title} {{Isotensional and
  isometric force-extension response of chains with bistable units and Ising
  interactions}},\ }\href {https://doi.org/10.1103/PhysRevE.98.052146}
  {\bibfield  {journal} {\bibinfo  {journal} {Physical Review E}\ }\textbf
  {\bibinfo {volume} {98}},\ \bibinfo {pages} {052146} (\bibinfo {year}
  {2018})}\BibitemShut {NoStop}%
\bibitem [{\citenamefont {Buche}\ and\ \citenamefont
  {Silberstein}(2020)}]{buche2020statistical}%
  \BibitemOpen
  \bibfield  {author} {\bibinfo {author} {\bibfnamefont {M.~R.}\ \bibnamefont
  {Buche}}\ and\ \bibinfo {author} {\bibfnamefont {M.~N.}\ \bibnamefont
  {Silberstein}},\ }\bibfield  {title} {\bibinfo {title} {{Statistical
  mechanical constitutive theory of polymer networks: The inextricable links
  between distribution, behavior, and ensemble}},\ }\href
  {https://doi.org/10.1103/PhysRevE.102.012501} {\bibfield  {journal} {\bibinfo
   {journal} {Physical Review E}\ }\textbf {\bibinfo {volume} {102}},\ \bibinfo
  {pages} {012501} (\bibinfo {year} {2020})}\BibitemShut {NoStop}%
\bibitem [{\citenamefont {Benedito}\ \emph {et~al.}(2020)\citenamefont
  {Benedito}, \citenamefont {Manca}, \citenamefont {Palla},\ and\ \citenamefont
  {Giordano}}]{benedito2020rate}%
  \BibitemOpen
  \bibfield  {author} {\bibinfo {author} {\bibfnamefont {M.}~\bibnamefont
  {Benedito}}, \bibinfo {author} {\bibfnamefont {F.}~\bibnamefont {Manca}},
  \bibinfo {author} {\bibfnamefont {P.~L.}\ \bibnamefont {Palla}},\ and\
  \bibinfo {author} {\bibfnamefont {S.}~\bibnamefont {Giordano}},\ }\bibfield
  {title} {\bibinfo {title} {Rate-dependent force-extension models for
  single-molecule force spectroscopy experiments},\ }\href
  {https://doi.org/10.1088/1478-3975/ab97a8} {\bibfield  {journal} {\bibinfo
  {journal} {Physical Biology}\ }\textbf {\bibinfo {volume} {17}},\ \bibinfo
  {pages} {056002} (\bibinfo {year} {2020})}\BibitemShut {NoStop}%
\bibitem [{\citenamefont {Florio}\ and\ \citenamefont
  {Puglisi}(2019)}]{florio2019unveiling}%
  \BibitemOpen
  \bibfield  {author} {\bibinfo {author} {\bibfnamefont {G.}~\bibnamefont
  {Florio}}\ and\ \bibinfo {author} {\bibfnamefont {G.}~\bibnamefont
  {Puglisi}},\ }\bibfield  {title} {\bibinfo {title} {Unveiling the influence
  of device stiffness in single macromolecule unfolding},\ }\href
  {https://doi.org/10.1038/s41598-019-41330-x} {\bibfield  {journal} {\bibinfo
  {journal} {Scientific Reports}\ }\textbf {\bibinfo {volume} {9}},\ \bibinfo
  {pages} {4997} (\bibinfo {year} {2019})}\BibitemShut {NoStop}%
\bibitem [{\citenamefont {Bellino}\ \emph {et~al.}(2019)\citenamefont
  {Bellino}, \citenamefont {Florio},\ and\ \citenamefont
  {Puglisi}}]{bellino2019influence}%
  \BibitemOpen
  \bibfield  {author} {\bibinfo {author} {\bibfnamefont {L.}~\bibnamefont
  {Bellino}}, \bibinfo {author} {\bibfnamefont {G.}~\bibnamefont {Florio}},\
  and\ \bibinfo {author} {\bibfnamefont {G.}~\bibnamefont {Puglisi}},\
  }\bibfield  {title} {\bibinfo {title} {The influence of device handles in
  single-molecule experiments},\ }\href {https://doi.org/10.1039/C9SM01376H}
  {\bibfield  {journal} {\bibinfo  {journal} {Soft Matter}\ }\textbf {\bibinfo
  {volume} {15}},\ \bibinfo {pages} {8680} (\bibinfo {year}
  {2019})}\BibitemShut {NoStop}%
\bibitem [{\citenamefont {Kreuzer}\ \emph {et~al.}(2001)\citenamefont
  {Kreuzer}, \citenamefont {Payne},\ and\ \citenamefont
  {Livadaru}}]{kreuzer2001stretching}%
  \BibitemOpen
  \bibfield  {author} {\bibinfo {author} {\bibfnamefont {H.}~\bibnamefont
  {Kreuzer}}, \bibinfo {author} {\bibfnamefont {S.}~\bibnamefont {Payne}},\
  and\ \bibinfo {author} {\bibfnamefont {L.}~\bibnamefont {Livadaru}},\
  }\bibfield  {title} {\bibinfo {title} {{Stretching a macromolecule in an
  atomic force microscope: Statistical mechanical analysis}},\ }\href
  {https://doi.org/10.1103/PhysRevE.63.021906} {\bibfield  {journal} {\bibinfo
  {journal} {Biophysical Journal}\ }\textbf {\bibinfo {volume} {80}},\ \bibinfo
  {pages} {2505} (\bibinfo {year} {2001})}\BibitemShut {NoStop}%
\bibitem [{\citenamefont {Buche}(2021)}]{buche2021fundamental}%
  \BibitemOpen
  \bibfield  {author} {\bibinfo {author} {\bibfnamefont {M.~R.}\ \bibnamefont
  {Buche}},\ }\emph {\bibinfo {title} {Fundamental Theories for the Mechanics
  of Polymer Chains and Networks}},\ \href {https://doi.org/10.7298/th3r-n996}
  {Ph.D. thesis},\ \bibinfo  {school} {Cornell University} (\bibinfo {year}
  {2021})\BibitemShut {NoStop}%
\bibitem [{\citenamefont {Buche}\ and\ \citenamefont
  {Silberstein}(2021)}]{buche2021chain}%
  \BibitemOpen
  \bibfield  {author} {\bibinfo {author} {\bibfnamefont {M.~R.}\ \bibnamefont
  {Buche}}\ and\ \bibinfo {author} {\bibfnamefont {M.~N.}\ \bibnamefont
  {Silberstein}},\ }\bibfield  {title} {\bibinfo {title} {Chain breaking in the
  statistical mechanical constitutive theory of polymer networks},\ }\href
  {https://doi.org/10.1016/j.jmps.2021.104593} {\bibfield  {journal} {\bibinfo
  {journal} {Journal of the Mechanics and Physics of Solids}\ }\textbf
  {\bibinfo {volume} {156}},\ \bibinfo {pages} {104593} (\bibinfo {year}
  {2021})}\BibitemShut {NoStop}%
\bibitem [{\citenamefont {Buche}\ \emph {et~al.}(2022)\citenamefont {Buche},
  \citenamefont {Silberstein},\ and\ \citenamefont
  {Grutzik}}]{buche2022freely}%
  \BibitemOpen
  \bibfield  {author} {\bibinfo {author} {\bibfnamefont {M.~R.}\ \bibnamefont
  {Buche}}, \bibinfo {author} {\bibfnamefont {M.~N.}\ \bibnamefont
  {Silberstein}},\ and\ \bibinfo {author} {\bibfnamefont {S.~J.}\ \bibnamefont
  {Grutzik}},\ }\bibfield  {title} {\bibinfo {title} {Freely jointed chain
  models with extensible links},\ }\href
  {https://doi.org/10.1103/PhysRevE.106.024502} {\bibfield  {journal} {\bibinfo
   {journal} {Physical Review E}\ }\textbf {\bibinfo {volume} {106}},\ \bibinfo
  {pages} {024502} (\bibinfo {year} {2022})}\BibitemShut {NoStop}%
\bibitem [{\citenamefont {Mulderrig}\ \emph {et~al.}(2023)\citenamefont
  {Mulderrig}, \citenamefont {Talamini},\ and\ \citenamefont
  {Bouklas}}]{mulderrig2023statistical}%
  \BibitemOpen
  \bibfield  {author} {\bibinfo {author} {\bibfnamefont {J.}~\bibnamefont
  {Mulderrig}}, \bibinfo {author} {\bibfnamefont {B.}~\bibnamefont
  {Talamini}},\ and\ \bibinfo {author} {\bibfnamefont {N.}~\bibnamefont
  {Bouklas}},\ }\bibfield  {title} {\bibinfo {title} {A statistical mechanics
  framework for polymer chain scission, based on the concepts of distorted bond
  potential and asymptotic matching},\ }\href
  {https://doi.org/10.1016/j.jmps.2023.105244} {\bibfield  {journal} {\bibinfo
  {journal} {Journal of the Mechanics and Physics of Solids}\ }\textbf
  {\bibinfo {volume} {174}},\ \bibinfo {pages} {105244} (\bibinfo {year}
  {2023})}\BibitemShut {NoStop}%
\bibitem [{\citenamefont {Buche}\ and\ \citenamefont
  {Grutzik}(2024)}]{buche2024statistical}%
  \BibitemOpen
  \bibfield  {author} {\bibinfo {author} {\bibfnamefont {M.~R.}\ \bibnamefont
  {Buche}}\ and\ \bibinfo {author} {\bibfnamefont {S.~J.}\ \bibnamefont
  {Grutzik}},\ }\bibfield  {title} {\bibinfo {title} {Statistical mechanical
  model for crack growth},\ }\href
  {https://doi.org/10.1103/PhysRevE.109.015001} {\bibfield  {journal} {\bibinfo
   {journal} {Physical Review E}\ }\textbf {\bibinfo {volume} {109}},\ \bibinfo
  {pages} {015001} (\bibinfo {year} {2024})}\BibitemShut {NoStop}%
\bibitem [{\citenamefont {Zwanzig}(1954)}]{zwanzig1954high}%
  \BibitemOpen
  \bibfield  {author} {\bibinfo {author} {\bibfnamefont {R.~W.}\ \bibnamefont
  {Zwanzig}},\ }\bibfield  {title} {\bibinfo {title} {{High-temperature
  equation of state by a perturbation method. I. Nonpolar gases}},\ }\href
  {https://doi.org/10.1063/1.1740409} {\bibfield  {journal} {\bibinfo
  {journal} {The Journal of Chemical Physics}\ }\textbf {\bibinfo {volume}
  {22}},\ \bibinfo {pages} {1420} (\bibinfo {year} {1954})}\BibitemShut
  {NoStop}%
\bibitem [{\citenamefont {McQuarrie}(2000)}]{mcq}%
  \BibitemOpen
  \bibfield  {author} {\bibinfo {author} {\bibfnamefont {D.~A.}\ \bibnamefont
  {McQuarrie}},\ }\href
  {https://uscibooks.aip.org/books/statistical-mechanics/} {\emph {\bibinfo
  {title} {Statistical Mechanics}}}\ (\bibinfo  {publisher} {University Science
  Books},\ \bibinfo {year} {2000})\BibitemShut {NoStop}%
\bibitem [{\citenamefont {Treloar}(1949)}]{treloar1949physics}%
  \BibitemOpen
  \bibfield  {author} {\bibinfo {author} {\bibfnamefont {L.~R.~G.}\
  \bibnamefont {Treloar}},\ }\href
  {https://global.oup.com/academic/product/the-physics-of-rubber-elasticity-9780198570271}
  {\emph {\bibinfo {title} {The Physics of Rubber Elasticity}}}\ (\bibinfo
  {publisher} {Clarendon Press},\ \bibinfo {year} {1949})\BibitemShut {NoStop}%
\bibitem [{\citenamefont {Flory}(1969)}]{flory1969statistical}%
  \BibitemOpen
  \bibfield  {author} {\bibinfo {author} {\bibfnamefont {P.~J.}\ \bibnamefont
  {Flory}},\ }\href
  {https://scholar.google.com/scholar_lookup?title=Statistical%20Mechanics%20of%20Chain%20Molecules&author=P.J.%20Flory&publication_year=1969}
  {\emph {\bibinfo {title} {Statistical Mechanics of Chain Molecules}}}\
  (\bibinfo  {publisher} {Interscience},\ \bibinfo {year} {1969})\BibitemShut
  {NoStop}%
\bibitem [{\citenamefont {Rubinstein}\ and\ \citenamefont
  {Colby}(2003)}]{rubinstein2003polymer}%
  \BibitemOpen
  \bibfield  {author} {\bibinfo {author} {\bibfnamefont {M.}~\bibnamefont
  {Rubinstein}}\ and\ \bibinfo {author} {\bibfnamefont {R.~H.}\ \bibnamefont
  {Colby}},\ }\href
  {https://global.oup.com/academic/product/polymer-physics-9780198520597}
  {\emph {\bibinfo {title} {Polymer Physics}}}\ (\bibinfo  {publisher} {Oxford
  University Press},\ \bibinfo {year} {2003})\BibitemShut {NoStop}%
\bibitem [{\citenamefont {Buche}(2023)}]{polymers}%
  \BibitemOpen
  \bibfield  {author} {\bibinfo {author} {\bibfnamefont {M.~R.}\ \bibnamefont
  {Buche}},\ }\href {https://doi.org/10.5281/zenodo.7041983} {\bibinfo {title}
  {\texttt{Polymers Modeling Library}}},\ \bibinfo {howpublished} {Zenodo}
  (\bibinfo {year} {2023}),\ \bibinfo {note}
  {\href{10.5281/zenodo.8066700}{v0.3.7}. {API} available in \texttt{Rust},
  \texttt{Python}, and \texttt{Julia}}\BibitemShut {NoStop}%
\bibitem [{\citenamefont {Thompson}\ \emph {et~al.}(2022)\citenamefont
  {Thompson}, \citenamefont {Aktulga}, \citenamefont {Berger}, \citenamefont
  {Bolintineanu}, \citenamefont {Brown}, \citenamefont {Crozier}, \citenamefont
  {in~'t Veld}, \citenamefont {Kohlmeyer}, \citenamefont {Moore}, \citenamefont
  {Nguyen}, \citenamefont {Shan}, \citenamefont {Stevens}, \citenamefont
  {Tranchida}, \citenamefont {Trott},\ and\ \citenamefont {Plimpton}}]{lammps}%
  \BibitemOpen
  \bibfield  {author} {\bibinfo {author} {\bibfnamefont {A.~P.}\ \bibnamefont
  {Thompson}}, \bibinfo {author} {\bibfnamefont {H.~M.}\ \bibnamefont
  {Aktulga}}, \bibinfo {author} {\bibfnamefont {R.}~\bibnamefont {Berger}},
  \bibinfo {author} {\bibfnamefont {D.~S.}\ \bibnamefont {Bolintineanu}},
  \bibinfo {author} {\bibfnamefont {W.~M.}\ \bibnamefont {Brown}}, \bibinfo
  {author} {\bibfnamefont {P.~S.}\ \bibnamefont {Crozier}}, \bibinfo {author}
  {\bibfnamefont {P.~J.}\ \bibnamefont {in~'t Veld}}, \bibinfo {author}
  {\bibfnamefont {A.}~\bibnamefont {Kohlmeyer}}, \bibinfo {author}
  {\bibfnamefont {S.~G.}\ \bibnamefont {Moore}}, \bibinfo {author}
  {\bibfnamefont {T.~D.}\ \bibnamefont {Nguyen}}, \bibinfo {author}
  {\bibfnamefont {R.}~\bibnamefont {Shan}}, \bibinfo {author} {\bibfnamefont
  {M.~J.}\ \bibnamefont {Stevens}}, \bibinfo {author} {\bibfnamefont
  {J.}~\bibnamefont {Tranchida}}, \bibinfo {author} {\bibfnamefont
  {C.}~\bibnamefont {Trott}},\ and\ \bibinfo {author} {\bibfnamefont {S.~J.}\
  \bibnamefont {Plimpton}},\ }\bibfield  {title} {\bibinfo {title}
  {{\texttt{LAMMPS} - a flexible simulation tool for particle-based materials
  modeling at the atomic, meso, and continuum scales}},\ }\href
  {https://doi.org/10.1016/j.cpc.2021.108171} {\bibfield  {journal} {\bibinfo
  {journal} {Computer Physics Communications}\ }\textbf {\bibinfo {volume}
  {271}},\ \bibinfo {pages} {108171} (\bibinfo {year} {2022})}\BibitemShut
  {NoStop}%
\bibitem [{\citenamefont {Mueller}\ \emph {et~al.}(2010)\citenamefont
  {Mueller}, \citenamefont {Van~Duin},\ and\ \citenamefont
  {Goddard~III}}]{mueller2010application}%
  \BibitemOpen
  \bibfield  {author} {\bibinfo {author} {\bibfnamefont {J.~E.}\ \bibnamefont
  {Mueller}}, \bibinfo {author} {\bibfnamefont {A.~C.}\ \bibnamefont
  {Van~Duin}},\ and\ \bibinfo {author} {\bibfnamefont {W.~A.}\ \bibnamefont
  {Goddard~III}},\ }\bibfield  {title} {\bibinfo {title} {Application of the
  \texttt{ReaxFF} reactive force field to reactive dynamics of hydrocarbon
  chemisorption and decomposition},\ }\href {https://doi.org/10.1021/jp9089003}
  {\bibfield  {journal} {\bibinfo  {journal} {The Journal of Physical Chemistry
  C}\ }\textbf {\bibinfo {volume} {114}},\ \bibinfo {pages} {5675} (\bibinfo
  {year} {2010})}\BibitemShut {NoStop}%
\end{thebibliography}%

\clearpage

\pagebreak
\widetext
\begin{center}%
    \textbf{\large%
        Supporting Information
        \\ \smallskip
        Modeling single-molecule stretching experiments using statistical thermodynamics
    }
    \\ \bigskip
    Michael R. Buche%
    \:\orcidlink{0000-0003-1892-0502}\,$^{1,}$\footnote{\href{mailto:mrbuche@sandia.gov}{mrbuche@sandia.gov}}
    and
    Jessica M. Rimsza%
    \:\orcidlink{0000-0003-0492-852X}\,$^2$
    \\ \smallskip
    {\small%
        \textit{
            {}$^1$Computational Solid Mechanics and Structural Dynamics,
            \\
            Sandia National Laboratories, Albuquerque, NM 87185, USA
            \\
            {}$^2$Geochemistry, Sandia National Laboratories, Albuquerque, NM 87185, USA
        }
        \\
        (Dated: \today)
    }
\end{center}

\setcounter{equation}{0}
\setcounter{figure}{0}
\setcounter{table}{0}
\setcounter{section}{0}
\makeatletter

\renewcommand{\theequation}{S\arabic{equation}}
\renewcommand{\thefigure}{S\arabic{figure}}
\renewcommand{\thesection}{S\roman{section}}
\renewcommand{\bibnumfmt}[1]{[S#1]}
\renewcommand{\citenumfont}[1]{S#1}

\section{Force-field parameters}

Below are the force-field parameters for the molecular dynamics calculations described in the manuscript:

{\footnotesize
\begin{verbatim}
Reactive MD-force field: Ni/C/H April 2009
39       ! Number of general parameters
    50.0000 !p(boc1)
    9.5469 !p(boc2)
    26.5405 !p(coa2)
    1.7224 !p(trip4)
    6.8702 !p(trip3)
    60.4850 !kc2
    1.0588 !p(ovun6)
    4.6000 !p(trip2)
    12.1176 !p(ovun6)
    13.3056 !p(ovun8)
    -70.5044 !p(trip1)
    0.0000 !Lower Taper-radius (swa)
    10.0000 !Upper Taper-radius (swb)
    2.8793 !Not used
    33.8667 !p(val7)
    6.0891 !p(lp1)
    1.0563 !p(val9)
    2.0384 !p(val10)
    6.1431 !Not used
    6.9290 !p(pen2)
    0.3842 !p(pen3)
    2.9294 !p(pen4)
    -2.4837 !Not used
    5.7796 !p(tor2)
    10.0000 !p(tor3)
    1.9487 !p(tor4)
    -1.2327 !not used
    2.1645 !p(cot2)
    1.5591 !p(vdW1)
    0.1000 !Cutoff for bond order*100  (cutoff)
    2.1365 !p(coa4)
    0.6991 !p(ovun4)
    50.0000 !p(ovun3)
    1.8512 !p(val8)
    0.5000 !Not used
    20.0000 !Not used
    5.0000 !not used
    0.0000 !not used
    2.6962 !p(coa3)
    3    ! Nr of atoms; atomID;ro(sigma); Val;atom mass;Rvdw;Dij;gamma;ro(pi);Val(e)
            alfa;gamma(w);Val(angle);p(ovun5);n.u.;chiEEM;etaEEM;n.u.
            ro(pipi);p(lp2);Heat increment;p(boc4);p(boc3);p(boc5),n.u.;n.u.
            p(ovun2);p(val3);n.u.;Val(boc);p(val5);n.u.;n.u.;n.u.
C    1.3831   4.0000  12.0000   1.8814   0.1923   0.9000   1.1363   4.0000
        9.7821   2.1317   4.0000  30.0000  79.5548   5.9666   7.0000   0.0000
        1.2071   0.0000 186.1720   9.0068  34.9357  13.5366   0.8563   0.0000
    -2.8983   2.5675   1.0564   4.0000   2.9663   0.0000   0.0000   0.0000
H    0.8873   1.0000   1.0080   1.5420   0.0598   0.6883  -0.1000   1.0000
        8.1910  30.9706   1.0000   0.0000 121.1250   3.5768  10.5896   1.0000
    -0.1000   0.0000  54.0596   1.3986   2.1457   0.0003   1.0698   0.0000
    -15.7683   2.1488   1.0338   1.0000   2.8793   0.0000   0.0000   0.0000
Ni   1.8201   2.0000  58.6900   1.9449   0.1880   0.8218   0.1000   2.0000
    12.1594   3.8387   2.0000   0.0000   0.0000   4.8038   7.3852   0.0000
    -1.0000   0.0000  95.6300  50.6786   0.6762   0.0981   0.8563   0.0000
    -3.7733   3.6035   1.0338   8.0000   2.5791   0.0000   0.0000   0.0000
    6      ! Nr of bonds; at1;at2;De(sigma);De(pi);De(pipi);p(be1);p(bo5);13corr;n.u.;p(bo6),p(ovun1)
                        p(be2);p(bo3);p(bo4);n.u.;p(bo1);p(bo2)
    1  1 143.3883  96.3926  76.4404  -0.7767  -0.4710   1.0000  34.9900   0.5108
        0.4271  -0.1116   9.0638   1.0000  -0.0840   6.7452   1.0000   0.0000
    1  2 181.9084   0.0000   0.0000  -0.4768   0.0000   1.0000   6.0000   0.7499
        12.8085   1.0000   0.0000   1.0000  -0.0608   6.9928   0.0000   0.0000
    2  2 168.2342   0.0000   0.0000  -0.2191   0.0000   1.0000   6.0000   1.0062
        6.1152   1.0000   0.0000   1.0000  -0.0889   6.0000   0.0000   0.0000
    1  3  83.5810   9.0383   0.0000   0.2531  -0.2000   1.0000  16.0000   0.0529
        1.4085  -0.1113  13.3900   1.0000  -0.1436   4.5683   1.0000   0.0000
    2  3 114.7566   0.0000   0.0000  -0.8939   0.0000   1.0000   6.0000   0.1256
        0.1054   1.0000   0.0000   1.0000  -0.1196   5.0815   0.0000   0.0000
    3  3  91.2220   0.0000   0.0000  -0.2538  -0.2000   0.0000  16.0000   0.2688
        1.4651  -0.2000  15.0000   1.0000  -0.1435   4.3908   0.0000   0.0000
    3    ! Nr of off-diagonal terms. at1;at2;Dij;RvdW;alfa;ro(sigma);ro(pi);ro(pipi)
    1  2   0.1188   1.4017   9.8545   1.1203  -1.0000  -1.0000
    1  3   0.0800   1.7085  10.0895   1.5504   1.4005  -1.0000
    2  3   0.0366   1.7306  11.1019   1.2270  -1.0000  -1.0000
18    ! Nr of angles. at1;at2;at3;Thetao,o;p(val1);p(val2);p(coa1);p(val7);p(pen1);p(val4)
    1  1  1  72.7917  38.5829   0.7209   0.0000   0.1409  17.4509   1.0670
    1  1  2  72.1533  14.2108   6.2512   0.0000   0.0100   0.0000   1.1022
    2  1  2  73.2608  24.9703   3.7807   0.0000   0.1335   0.0000   3.0461
    1  2  2   0.0000   0.0000   6.0000   0.0000   0.0000   0.0000   1.0400
    1  2  1   0.0000   7.5000   5.0000   0.0000   0.0000   0.0000   1.0400
    2  2  2   0.0000  27.9213   5.8635   0.0000   0.0000   0.0000   1.0400
    1  3  1  62.5000  16.6806   0.7981   0.0000   0.9630   0.0000   1.0711
    1  1  3  87.6241  12.6504   1.8145   0.0000   0.6154   0.0000   1.5298
    3  1  3 100.0000  40.4895   1.6455   0.0000   0.0100   0.0000   1.7667
    1  3  3   5.0994   3.1824   0.7016   0.0000   0.7465   0.0000   2.2665
    2  3  2 106.3969  30.0000   0.9614   0.0000   1.9664   0.0000   2.2693
    2  2  3   0.0000  26.3327   4.6867   0.0000   0.8177   0.0000   1.0404
    3  2  3   0.0000  60.0000   1.8471   0.0000   0.6331   0.0000   1.8931
    2  3  3  30.3748   1.0000   4.8528   0.0000   0.1019   0.0000   3.1660
    2  3  3 180.0000 -27.2489   8.3752   0.0000   0.8112   0.0000   1.0004
    1  3  2  97.5742  10.9373   2.5200   0.0000   1.8558   0.0000   1.0000
    1  2  3   0.0000   0.2811   1.1741   0.0000   0.9136   0.0000   3.8138
    2  1  3  84.0006  45.0000   0.6271   0.0000   3.0000   0.0000   1.0000
12    ! Nr of torsions. at1;at2;at3;at4;;V1;V2;V3;p(tor1);p(cot1);n.u;n.u.
    1  1  1  1  -0.5000  53.0886  -0.1335  -6.2875  -1.9524   0.0000   0.0000
    1  1  1  2  -0.4614  29.0459   0.2551  -4.8555  -2.7007   0.0000   0.0000
    2  1  1  2  -0.2833  31.2867   0.2965  -4.8828  -2.4652   0.0000   0.0000
    0  1  2  0   0.0000   0.0000   0.0000   0.0000   0.0000   0.0000   0.0000
    0  2  2  0   0.0000   0.0000   0.0000   0.0000   0.0000   0.0000   0.0000
    0  1  1  0   0.0000  50.0000   0.3000  -4.0000  -2.0000   0.0000   0.0000
    1  1  1  3   0.0000   5.0000   0.4000  -6.0000   0.0000   0.0000   0.0000
    3  1  1  3   0.0000  44.3024   0.4000  -4.0000   0.0000   0.0000   0.0000
    2  1  1  3   0.0000  21.7038   0.0100  -4.0000   0.0000   0.0000   0.0000
    2  1  3  1   0.0000   5.2500   0.0100  -6.0000   0.0000   0.0000   0.0000
    1  1  3  1   0.0000   5.1676   0.0100  -5.9539   0.0000   0.0000   0.0000
    1  1  3  2   0.0000   5.1676   0.0100  -5.9539   0.0000   0.0000   0.0000
    0    ! Nr of hydrogen bonds. at1;at2;at3;r(hb);p(hb1);p(hb2);p(hb3)
\end{verbatim}
}

\end{document}